\documentclass[12pt]{article}

\begin{document}
\input amssym.tex


\title{Canonical quantization of the covariant fields on de Sitter spacetimes}

\author{ION I. COTAESCU \footnote{e-mail: i.cotaescu@e-uvt.ro}\\{\it West University of Timi\c soara, V. Parvan Ave. 4,}\\
{\it Timi\c soara, 300223, Romania }}

\maketitle


\begin{abstract}
The properties of the covariant quantum fields on  de Sitter spacetimes are investigated focusing on the isometry generators and Casimir operators in order to establish the equivalence among the covariant representations and the unitary irreducible ones of the de Sitter isometry group. For the Dirac quantum field it is shown that the spinor covariant representation, transforming the Dirac field under de Sitter isometries, is equivalent with a direct sum of two unitary irreducible representations of the $Sp(2,2)$ group, transforming alike the particle and antiparticle field operators in momentum representation. Their basis generators and Casimir operators are written down finding that the covariant representations are equivalent with unitary irreducible ones from the principal series whose canonical  weights are determined by the fermion mass and spin.

Pacs: 04.62.+v
\end{abstract}

\newpage

\tableofcontents

\newpage
\section{Introduction}

The principal invariant quantities determining the nature of the elementary particles are the mass and spin. In the quantum theory of free fields the mass and spin are defined exclusively by the invariants of the spacetime symmetries in the sense that these are encapsulated in the eigenvalues of the Casimir operators of the unitary irreducible representations (UIRs) of the isometry group. Moreover, the interactions that are governed by other symmetries, the internal and gauge ones, cannot affect the meaning of these geometric invariants even though the mass has to be redefined in the renormalization procedure. 

We have shown that the covariant fields of  general relativity, defined on the $(1+3)$-dimensional curved spacetimes, $(M,g)$, transform under isometries according to covariant representations (CRs)  {\em induced} by finite-dimensional representations (reps.) of the universal covering group $\hat G$ of the gauge one, $G$ \cite{ES,ES1}. In the case of  local-Minkowskian manifolds, under consideration here, these groups are $G=SO(1,3)$ and respectively $\hat G=SL(2,{\Bbb C})$.  For this reason, the spin terms of the operators generating CRs are given by linear reps. of the $sl(2,{\Bbb C})$ algebra instead of the  $s(M)$ algebra of the universal covering group $S(M)$ of the isometry one, $I(M)$.  

On the other hand, the induced CRs must be equivalent with orthogonal sums of UIRs of the  group $S(M)$.  In special relativity, the CRs are equivalent with orthogonal sums of  Wigner's UIRs  that govern the transformation rules under isometries of the particle and antiparticle operators in momentum rep. \cite{Wigner, Mck,WKT,BARA}. This result is related to the special structure of the  Poincar\' e  isometry group, $T(4)\circledS G$, which cannot be generalized to other manifolds, not even in the case of the de Sitter spacetime  which still allows a momentum rep..  Nevertheless, the CRs of the de Sitter external symmetry group must be related to the UIRs of the same group which are well-known \cite{linrep,linrep1}. In what follows we shall pay a special attention to this problem we refer hare as the CR-UIR equivalence.

The de Sitter manifold, denoted from now by $M$, is local-Minkowskian and has the isometry group $I(M)=SO(1,4)$ which is the gauge group of the Minkowskian five-dimensional manifold $M^5$ embedding $M$. The UIRs of the corresponding group  $S(M)={\rm Spin}(1,4)=Sp(2,2)$  \cite{linrep,linrep1} are used in various applications. Many authors exploited this high symmetry for building quantum theories,  either by constructing symmetric two-point functions, avoiding thus the canonical quantization \cite{Wood,Wood1}, or by using directly these UIRs for writing field equations without considering  CRs \cite{Gaz,Gaz1,Gaz2}. A different approach, which applies the {\em canonical quantization} to the covariant fields transforming according to {\em induced} CRs, was initiated by Nachtmann \cite{Nach}  many years ago and continued in few of our papers \cite{CNach,CPF,CSB,CCC} where we developed our theory of induced CRs, creating thus the framework of the de Sitter QED in Coulomb gauge \cite{CQED}. 

In this  report we would like to present this approach focusing on our principal results concerning the properties of the CRs and UIRs of the de Sitter isomety group, the CR-UIR equivalence  and the role of the generators of these reps.  in  determining the principal invariants of the quantum field theory (QFT) on the de Sitter spacetime \cite{CCC,Cnew}. 

On this manifold we cannot apply the Wigner method for investigating the structure of the covariant fields such that we must study the general features of the CRs in the configuration rep., deriving the principal invariants. In this manner, we find the form of  the generators of  the induced CRs and the corresponding Casimir operators that help us to establish indirectly the equivalence of the CRs with orthogonal sums of UIRs of the group $S(M)$ \cite{CCC}. However, at this level, we cannot deduce the form of the UIRs generators  in momentum rep.  which act on the particle and antiparticle field operators.  This is because   the absence of a general Wigner theory that forces us to use field equations for determining the structure of the covariant fields in each particular case separately.     

For this reason, we concentrate on the Dirac field, investigating the relation among the CRs and UIRs of the Dirac theory in momentum rep. on the de Sitter background \cite{Cnew}.  As mentioned,  these CRs  are induced by the linear reps. of the group $\hat G$ without to meet explicitly  the linear reps. of the group $S(M)$.  Nevertheless,  the equivalence between a Dirac CR and a pair of the UIRs of the group $S(M)$ can be proved by constructing the UIR generators in momentum rep. giving rise, after quantization, to the conserved one-particle operators of the QFT including the Casimir ones.   In this manner we find that the Dirac particle and antiparticle operators in momentum rep. transform according to the {\em same} UIR that can be one of the equivalent UIRs $(s,q)$,  from the principal series \cite{linrep,linrep1},  labeled by  the spin $s=\frac{1}{2}$ and $q=\frac{1}{2}\pm  i\frac{m}{\omega}$ where $m$ is the fermion mass and $\omega$ is the  Hubble constant of $M$  in our notation. Note that the fundamental spinors we use here correspond to a fixed vacuum of the Bounch-Davies type \cite{BD}  as in our de Sitter QED \cite{CQED}. 

This result is similar with that of special relativity where the particle and antiparticle operators in momentum rep. of any covariant quantum field with unique spin, $s$, transform alike under Poincar\'e isometries, according to the same Wigner UIR induced by the $(2s+1)$-dimensional UIR  of the group $SU(2)$ \cite{W}. Obviously, this happens only if we respect the  connection between spin and statistic, assuming that  the Dirac particle and antiparticle operators satisfy canonical anti-commutation rules. 

On the other hand, the concrete forms of generators presented here  allow us to expand  in momentum rep.  the conserved one-particle operators corresponding to the de Sitter isometries via Noether theorem. We show that for all these operators (energy, momentum, angular momentum, etc.) the contributions of the particles and antiparticles are {\em additive} in contrast with the conserved charge where these have opposite signs.  We find thus that an important feature of the QFT on Minkowski spacetime can be retrieved on curved backgrounds.

This paper is organized as follows. In the next section we present our general theory of covariant fields on curved spacetimes giving the general form of the CRs and their generators. In the next section we revisit the method of canonical quantization in a Lagrangian QFT where we can use the mode expansions  in a given rep. determined by a set of commuting operators. The CRs of special relativity and the Wigner theory of the induced UIRs of the Poincar\' e group are discussed briefly in the fourth section giving as example the free Dirac field. The next section is devoted to the theory of covariant fields on de Sitter spacetime, presenting the general form of the generators of the CRs of the de Sitter isometries and the corresponding Casimir operators. In the sixth section we concentrate on the Dirac field on de Sitter spacetime  giving the fundamental solutions of the free Dirac equation and  deriving the generators of the UIRs in momentum rep. and the components of the Pauli-Lubanski operator which help us to obtain the Casimir operators.  Moreover, we give the general form of the momentum expansions of the principal conserved observables of  QFT discussing  their properties and the CR-UIR equivalence at this level. Finally, we present our concluding remarks.

\section{Covariant  fields}

The covariant fields of special relativity transform under the Poincar\' e isometries according to finite-dimensional reps. of the Larentz group  which is a mere subgroup of the isometry one. In what follows we shall see that this is not a contingency since in any curved spacetime the CRs are induced by the gauge  group of the flat metric of the pseudo-Euclidean model of this manifold. 

\subsection{Induced CRs}

Any  local-Minkowskian spacetime $(M,g)$  may be equipped with {\em local}  frames $\{x;e\}$ formed by a local chart (or natural frame) $\{x\}$ and a non-holonomic orthogonal frame $\{e\}$. In a given local chart of coordinates $x^{\mu}$, labelled by natural indices, $\mu, \nu,...=0,1,2,3$, the orthogonal frames and the corresponding coframes, $\{\hat e\}$, are defined by the tetrad fields $e_{\hat\mu}$  and  $\hat e^{\hat\mu}$, which are labelled by local indices, $\hat\mu, \hat\nu,...=0,1,2,3$, and obey the usual duality relations,  $\hat e^{\hat\mu}_{\alpha}\,
e_{\hat\nu}^{\alpha}=\delta^{\hat\mu}_{\hat\nu}$, $ \hat
e^{\hat\mu}_{\alpha}\, e_{\hat\mu}^{\beta}=\delta^{\beta}_{\alpha}$,
and the orthonormalization conditions,  $e_{\hat\mu}\cdot e_{\hat\nu}=\eta_{\hat\mu
\hat\nu}$, $\hat e^{\hat\mu}\cdot \hat e^{\hat\nu}=\eta^{\hat\mu
\hat\nu}$, given by the metric $\eta=$diag$(1,-1,-1,-1)$ of the Minkowski spacetime $(M_0,\eta)$ which is the pseudo-Euclidean model of $(M,g)$.  

The tetrad fields define the local derivatives $\hat\partial_{\hat\alpha}=e^{\mu}_{\hat\alpha}\partial_{\mu}$
and the basis 1-forms $\tilde\omega^{\hat\alpha}(x)=\hat e^{\hat\alpha}_{\mu}(x)dx^{\mu}$ giving the metric tensor $g_{\mu
\nu}=\eta_{\hat\alpha\hat\beta}\hat e^{\hat\alpha}_{\mu}\hat
e^{\hat\beta}_{\nu}$ that raises or lowers the natural indices while for
the local ones  we have to use the flat metric $\eta$.

The metric $\eta$ remains invariant under the transformations of the group $O(1,3)$  which includes the Lorentz group, $L_{+}^{\uparrow}$, whose universal covering group is $SL(2,\Bbb C)$.  In the usual covariant parametrization, with the real parameters, $\omega^{\hat\alpha \hat\beta}=-\omega^{\hat\beta\hat\alpha}$, the transformations 
\begin{equation}
A(\omega)=\exp\left(-\frac{i}{2}\omega^{\hat\alpha\hat\beta}
S_{\hat\alpha\hat\beta}\right) \in SL(2,\Bbb C) 
\end{equation}
depend on the covariant basis-generators of the $sl(2,{\Bbb C})$ Lie algebra, $S_{\hat\alpha\hat\beta}$, which satisfy the commutation rules
\begin{equation}\label{comSS}
{[}S_{\hat\mu\hat\nu},S_{\hat\sigma\hat\tau}{]}=i(
\eta_{\hat\mu\hat\tau}S_{\hat\nu\hat\sigma}-\eta_{\hat\mu\hat\sigma}S_{\hat\nu\hat\tau}+\eta_{\hat\nu\hat\sigma}S_{\hat\mu\hat\tau}
-\eta_{\hat\nu\hat\tau}S_{\hat\mu\hat\sigma})\,.
\end{equation} 
Bearing in mind that the $sl(2,{\Bbb C})$ algebra has two fundamental ireps., $(\frac{1}{2},0)$ and $(0.\frac{1}{2})$, which are non-equivalent, it is convenient to consider as fundamental rep. just the reducible spinor rep. $\rho_D=(\frac{1}{2},0)\oplus(0,\frac{1}{2})$ of the Dirac theory, denoting $\rho_D(A)=A\,, \forall A\in SL(2,{\Bbb C})$ and  $\rho_D(S)=S\,, \forall S\in sl(2,{\Bbb C})$. The advantage is that for any rep. of the Dirac matrices $\gamma^{\hat\alpha}$ we have
\begin{equation}\label{Sgg}
S^{\hat\alpha\hat\beta}=\frac{i}{4}\left[\gamma^{\hat\alpha},\gamma^{\hat\beta}\right]\,.
\end{equation}
More details are given in the  Appendices A and B.

The parametrization used here offers us, in addition, the advantage of a simple expansion of the matrix elements
in local bases,
$\Lambda^{\hat\mu\,\cdot}_{\cdot\,\hat\nu}[A(\omega)]=
\delta^{\hat\mu}_{\hat\nu}
+\omega^{\hat\mu\,\cdot}_{\cdot\,\hat\nu}+\cdots$, of the
transformations $\Lambda[A(\omega)]\in L_{+}^{\uparrow}$ associated
to $A(\omega)$ through the canonical homomorphism \cite{WKT}. When  $(M,g)$ is assumed to be orientable and time-orientable we may consider the Lorentz group, $G(\eta)=L^{\uparrow}_{+}$,  as the gauge group of the Minkowski metric $\eta$ \cite{Wald}.

The {\em covariant fields}, $\psi_{(\rho)}:\, M\to {\cal V}_{(\rho)}$, are locally defined over $M$ with values in the vector spaces ${\cal V}_{(\rho)}$ carrying the finite-dimensional non-unitary reps. $\rho$ of the group $SL(2,\Bbb C)$. We denote by ${\cal F}_{\rho}$ the space of these fields supposed to have 'good' properties (being  at least of  class $C^2$)  but which will be organized leter. In general, the reps. $\rho$ are reducible being equivalent with direct sums of irreducible reps. (irreps.), $(j_1,j_2)$ \cite{WKT,BARA} as we show in the Appendix A. They determine the form of the covariant derivatives of the field $\psi_{(\rho)}$ in local frames,
\begin{equation}\label{der}
D_{\hat\alpha}^{(\rho)}= e_{\hat\alpha}^{\mu}D_{\mu}^{(\rho)}=
\hat\partial_{\hat\alpha}+\frac{i}{2}\, \rho(S^{\hat\beta\, \cdot} _{\cdot\,
\hat\gamma})\,\hat\Gamma^{\hat\gamma}_{\hat\alpha \hat\beta}\,.
\end{equation}
which depend on the connection coefficients  in local frames
\begin{equation}
\hat\Gamma^{\hat\sigma}_{\hat\mu \hat\nu}=e_{\hat\mu}^{\alpha}
e_{\hat\nu}^{\beta}(\hat e_{\gamma}^{\hat\sigma}
\Gamma^{\gamma}_{\alpha \beta} -\hat e^{\hat\sigma}_{\beta,
\alpha})\,, 
\end{equation}
where $\Gamma^{\gamma}_{\alpha \beta}$ denote the Christoffel symbols.
These covariant derivative  assure the covariance of the whole theory under the (point-dependent) tetrad-gauge transformations, 
\begin{eqnarray}
\tilde\omega & \to &{\tilde\omega}'= \Lambda[A]\tilde\omega\,,\\
\psi_{(\rho)}&\to & \psi_{(\rho)}'=  \rho(A)\psi_{(\rho)}\,,\label{Gauge}
\end{eqnarray}
produced by the sections $A\in SL(2,{\Bbb C})$ of the spin fiber bundle \cite{Wald}. 

When $(M,g)$ has isometries, $x\to x'=\phi_{\frak g} (x)$, these are, generally, non-linear reps. ${\frak g}\to \phi_{\frak g}$ of the isometry group $I(M)$ complying with the composition rule $\phi_{\frak g}\circ \phi_{{\frak g}'}=\phi_{\frak{gg}'}$, $\forall \frak{g,g}'\,\in I(M)$. Denoting then by $id=\phi_{\frak e}$ the identity function, corresponding to the unit ${\frak e}\in I(M)$, we deduce $\phi_{\frak g}^{-1}=\phi_{{\frak g}^{-1}}$. In a given parametrization, ${\frak g}={\frak g}(\xi)$ (with ${\frak e}={\frak g}(0)$), the isometries
\begin{equation}
x\to x'=\phi_{{\frak g}(\xi)}(x)=x+\xi^a k_a(x) +... 
\end{equation}
lay out the Killing vectors $k_a=\partial_{\xi_a}\phi_{{\frak g}(\xi)}|_{\xi=0}$ associated to the parameters $\xi^a$ ($a,b,...=1,2...N$).  

In general, the isometries may change the relative position of the local frames affecting thus the physical interpretation. For this reason we proposed the theory of external symmetry \cite{ES} where we introduced the combined transformations $(A_{\frak g},\phi_{\frak g})$  able to correct the position of the local frames. These transformations must preserve not only the metric  but the tetrad-gauge too, transforming  the 1-forms as $\tilde\omega(x')=\Lambda[A_{\frak g}(x)]\tilde\omega(x)$. Hereby, we deduce  \cite{ES},
\begin{equation}\label{Axx}
\Lambda^{\hat\alpha\,\cdot}_{\cdot\,\hat\beta}[A_{\frak g}(x)]= \hat
e_{\mu}^{\hat\alpha}[\phi_{\frak g}(x)]\frac{\partial
\phi^{\mu}_{\frak g}(x)} {\partial x^{\nu}}\,e^{\nu}_{\hat\beta}(x)\,,
\end{equation}
assuming, in addition, that  $A_{{\frak g}={\frak e}}(x)=1\in SL(2,\Bbb C)$. 
Then, the combined 
transformations $(A_{\frak g},\phi_{\frak g})$ preserve the gauge,  
\begin{equation}
(A_{\frak g},\phi_{\frak g}):\quad
\begin{array}{rlrcl}
e(x)&\to&e'(x')&=&e[\phi_{\frak g}(x)]\,,\\
\hat e(x)&\to&\hat e'(x')&=&\hat e[\phi_{\frak g}(x)]\,,
\end{array}
\qquad
\end{equation}
transforming the covariant fields according to the rule
\begin{equation}
(A_{\frak g},\phi_{\frak g}):\quad \psi_{(\rho)}(x)\to\psi_{(\rho)}'(x')=\rho[A_{\frak g}(x)]\psi_{(\rho)}(x)\,,\label{es}
\end{equation}
which defines the operator-valued CR  $T^{(\rho)} \,:\, (A_{\frak g},\phi_{\frak g})\to T_{\frak g}^{(\rho)}$, induced by $\rho$,   whose operators act as
\begin{equation}\label{Tx}
(T_{\frak g}^{(\rho)}\psi_{(\rho)})[\phi_{\frak g}(x)]=\rho[A_{\frak g}(x)]\psi_{(\rho)}(x)\,.
\end{equation}
We have shown that the pairs $(A_{\frak g},\phi_{\frak g})$ constitute a well-defined Lie group with respect to the new operation  that can be seen as a rep. of the universal covering group of $I(M)$ denoted here by $S(M)$ \cite{ES}. 

\subsection{Conserved observables}

In a given parametrization, ${\frak g}={\frak g}(\xi)$, for small values of  
$\xi^{a}$, the parameters of $A_{{\frak g}(\xi)}(x)\equiv
A[\omega_{\xi}(x)]$ can be expanded as
$\omega^{\hat\alpha\hat\beta}_{\xi}(x)=
\xi^{a}\Omega^{\hat\alpha\hat\beta}_{a}(x) +\cdots$, in terms of the
functions
\begin{equation}\label{Om}
\Omega^{\hat\alpha\hat\beta}_{a}\equiv {\frac{\partial
\omega^{\hat\alpha\hat\beta}_{\xi}} {\partial\xi^a}}_{|\xi=0}
=\left( \hat e^{\hat\alpha}_{\mu}\,k_{a,\nu}^{\mu} +\hat
e^{\hat\alpha}_{\nu,\mu}
k_{a}^{\mu}\right)e^{\nu}_{\hat\lambda}\eta^{\hat\lambda\hat\beta}
\end{equation}
that  are skew-symmetric,
$\Omega^{\hat\alpha\hat\beta}_{a}=-\Omega^{\hat\beta\hat\alpha}_{a}$,
only if $k_a$ are Killing vectors \cite{ES}. In this case we deduce the  basis-generators
\cite{ES},
\begin{equation}\label{Xa}
X_{a}^{(\rho)}=i{\partial_{\xi^a} T_{{\frak g}(\xi)}^{(\rho)}}_{|\xi=0}=-i
k_a^{\mu}\partial_{\mu} +\frac{1}{2}\,\Omega^{\hat\alpha\hat\beta}_{a}
S_{\hat\alpha\hat\beta}^{(\rho)}\,,
\end{equation}
where we use the equivalent notation $S_{\hat\alpha\hat\beta}^{(\rho)}=\rho(S_{\hat\alpha\hat\beta})$. 
These operators   satisfy the commutation rules 
\begin{equation}\label{XaXb}
[X_{a}^{(\rho)}, X_{b}^{(\rho)}]=ic_{abc}X_{c}^{(\rho)}
\end{equation}
determined by the structure constants, $c_{abc}$, of the algebras $s(M)\sim i(M)$. In other words, they are the basis-generators of a CR of the $s(M)$ algebra {\em induced} by the rep. $\rho$ of the $sl(2,{\Bbb C})$ algebra.  Therefore, they transform under isometries ${\frak g}={\frak g}(\xi)$ according to the adjoint representation of $S(M)$,
\begin{equation}\label{TxX}
T_{\frak g}^{(\rho)}X^{(\rho)}_a{T_{\frak g}^{(\rho)}}^{-1}=Ad({\frak g})_{a b}X_b^{(\rho)}\,,
\end{equation}
defined canonically as \cite{WKT},
\begin{equation}
Ad({\frak g})=e^{-i\xi^a ad(X_a)}\,, \quad ad(X_a)_{bc}=ic_{abc}\,.
\end{equation}

The form of the generators (\ref{Xa}) is appearently non-covariant since this is written in local frames. However, in natural frames, they can be rewritten in an equivalent covariant form  \cite{ES1},
\begin{equation}\label{CM}
X^{(\rho)}_{a}=-ik^{\mu}_{a}D_{\mu}^{(\rho)}+\frac{1}{2}\, k_{a\,
\mu;\,\nu}\,e^{\mu}_{\hat\alpha}\,e^{\nu}_{\hat\beta}\,
\rho(S^{\hat\alpha\hat\beta})\,,
\end{equation}
which represents the generalization to any representation $\rho$ of the formula given by Carter and McLenaghan  for the Dirac field  \cite{CML}. Note that operators of this type are known  from long time being proportional with the  Kosmann Lie derivatives associated to the Killing vectors \cite{Kos}. Nevertheless, their role in generating CRs on any curved manifold having isometries was demonstrated for the first time just in our theory of external symmetry we present here \cite{ES}. 

The generators (\ref{Xa}) or (\ref{CM}) are the principal conserved observables of the quantum theory which commute with the operators of the field equations resulted from an invariant Lagrangian theory. These generators have, in general, point-dependent spin terms which do not commute with the orbital parts. However, there are tetrad-gauges in which at least the generators of a subgroup  $H \subset I(M)$ may have point-independent spin terms commuting with the orbital parts. Then we say that the restriction to $H$ of the CR $T^{(\rho)}$ is {\em manifest} covariant \cite{ES}. Obviously, if $H=I(M)$ then the whole
rep. $T^{(\rho)}$ is manifest covariant. In particular, the linear CRs on the Minkowski spacetime have this property. 

Finally, we must specify that the above definition of the CRs transforming the covariant fields is general, including all the particular cases studied so far. Thus the covariant fields with  integer spin which are independent on the local frames are just the vectors and tensors of any rank  \cite{ES}.  Any tensor field, $\Theta$, transforms under isometries as $\Theta\to \Theta'=T_{\frak g}\Theta$, according to a  tensor representation of the group $S(M)$
defined by the well-known rule in natural frames
\begin{equation}
\left[\frac{\partial \phi^{\alpha}_{\frak g}(x)}{\partial x_{\mu}}\frac{\partial
\phi^{\beta}_{\frak g}(x)}{\partial x_{\nu}}...
\right]\left(T_{\frak g}\Theta\right)_{\alpha\beta...}[\phi_{\frak g}(x)]=\Theta_{\mu\nu...}(x)\,,
\end{equation}
Hereby one derives the basis-generators of the tensor representation,
$X_{a}=i\,\partial_{\xi_a}T_{\frak g}|_{\xi=0}$, whose action, 
\begin{equation}\label{XT}
(X_a\,
\Theta)_{\alpha\beta...}=-i({k_a}^{\nu}\Theta_{\alpha\beta...;\,\nu}
+{k_a}^{\nu}_{~;\,\alpha}\Theta_{\nu\beta...}
+{k_a}^{\nu}_{~;\,\beta}\Theta_{\alpha\nu ...}...)\,,
\end{equation}
is the same as that of the operators (\ref{CM}) \cite{ES1}. Therefore, the CRs are useful especially in theories involving covariant fields  with half integer spin, depending explicitly  on the choice of the orthogonal local frames.    

\section{Quantum theory of covariant fields}

The generators of the CRs are the quantum observables of the relativistic quantum mechanics that are conserved in the sense that they commute with the operator of the field equation which is invariant under isometries. The next step is the second quantization that cannot be performed in a canonical manner without the framework of a Lagrangian field theory where the Noether theorem give rise to classical conserved quantities which become the conserved one-particle operators of the quantum theory. In our approach these operators will be just the generators of the CRs of quantum fields. 

\subsection{Lagrangian formalism}

The construction of the Lagrangian theory of covariant fields is based on some  positive defined quadratic forms which must remain  invariant under the action of the transformations $\rho(A)$. 

Since the finite-dimensional reps. $\rho$ of the $SL(2,{\Bbb C})$ group  are non-unitary,  we need to use reducible reps. and the (generalized) Dirac conjugation, $\overline\psi_{(\rho)}=\psi^+_{\rho} \gamma_{(\rho)}$, where the matrix $\gamma_{(\rho)}=\gamma^+_{(\rho)}=\gamma^{-1}_{(\rho)}$ must be chosen  such that 
\begin{equation}
\overline{\rho (A)}=\gamma_{(\rho)}\rho(A)^+\gamma_{(\rho)}=\rho(A^{-1})\,,
\end{equation}
Then the generators of the rep. $\rho$ are self-adjoint with respect to the Dirac conjugation, $\overline{\rho(S)}=\rho(S)$ and  the quadratic form $\overline\psi_{(\rho)} \psi_{(\rho)}$ is invariant under the gauge transformations (\ref{Gauge}). In general, the Dirac conjugation can be defined for the {\em symmetric} reps. $\rho$ which are direct sums including only self-adjoint irreps. and  pairs of adjoint irreps., $\rho=...(j,j)\oplus...(j_1,j_2)\oplus (j_2,j_1)...$, as we briefly argue in Appendix A.  In this manner, the spin content of the theory, denoted by ${\Bbb S}(\rho)$, is increasing since each irrep.  $ (j_1,j_2)$ brings the subspeces ${\cal V}_s$  of the UIRs of the group $SU(2)$ with spins  $s=j_1+j_2, j_1+j_2-1,...|j_1-j_2|$  \cite{WKT}. Then the carrier space of the rep. $\rho$ can be decomposed as
\begin{equation}
{\cal V}_{(\rho)}=\sum_{s\in {\Bbb S}(\rho)}\oplus {\cal V}_{\sigma}\,.
\end{equation}
We remind the reader that the irreducible reps. (irreps.) with unique spin $s$ are only $(s,0)$ and $(0,s)$. 

Under such circumstances it is difficult to construct covariant fields with {\em unique} spin $s$, eliminating the unwanted components.  The simplest method was proposed by Weinberg \cite{W} which assumed that a covariant field of spin $s$ transforms according to the CR induced by the rep. $\rho_s=(s,0)\oplus (0,s)$.  The typical example is the Dirac field  with $s=\frac{1}{2}$ and $\rho_D=(\frac{1}{2},0)\oplus(0,\frac{1}{2})$ which will be studied later. However, this method leads to field equations with derivatives up to the order $2s$ which is a serious impediment requiring Lagrangian theories with higher order derivatives.  An alternative method  was  proposed by Fronsdal in Minkowski spacetime \cite{Fr1,Fr2,Fr3} assuming that the covariant fields of integer spins transform according to the self-adjoint irreps. $(\frac{s}{2},\frac{s}{2})$ while for the spin half integer one must use  the reps.  $\rho=\rho_D \otimes (\frac{s}{2}-\frac{1}{4},\frac{s}{2}-\frac{1}{4})$. In this approach the uniqueness of the spin is assured by special field equations of first or second order having spin-dependent coefficients. Unfortunately this method works only for massless fields. Thus we conclude that the general problem of constructing Lagrangian theories of massive covariant fields with unique spin and field equations of at most second order derivatives is not yet solved.

The usual covariant  free fields satisfy field equations having only first or second order derivatives which can be derived from actions of the form
\begin{equation}
{\cal S}[\psi_{(\rho)},\overline{\psi}_{(\rho)}]=\int_{\Delta} d^4 x\,\sqrt{g}\, {\cal
L}(\psi_{(\rho)},\psi_{(\rho) ;\mu},\overline{\psi}_{(\rho)},\overline{\psi}_{(\rho);\mu})\,,\quad g=|{\rm
det}\, g_{\mu\nu}|\,,
\end{equation}
depending on the field $\psi_{(\rho)}$, its Dirac adjoint $\overline{\psi}_{(\rho)}$ and their corresponding covariant derivatives $\psi_{(\rho) ;\mu}=D_{\mu}\psi_{(\rho)}$ and $\overline{\psi}_{(\rho);\mu}=\overline{D_{\mu}\psi}_{(\rho)}$ defined by the rep. $\rho$ of the group $SL(2,{\Bbb C})$.    The action ${\cal S}$ is extremal if the covariant fields  satisfy the Euler-Lagrange equations
\begin{equation}
\frac{\partial {\cal L}}{\partial
\overline{\psi}_{(\rho)}}-\frac{1}{\sqrt{g}}\,\partial_{\mu}\,\frac{\partial
{(\sqrt{g}\,\cal L})}{\partial \overline{\psi}_{(\rho) ,\mu}}=0\,,\qquad
\frac{\partial {\cal L}}{\partial
{\psi}_{(\rho)}}-\frac{1}{\sqrt{g}}\,\partial_{\mu}\,\frac{\partial
{(\sqrt{g}\,\cal L})}{\partial {\psi}_{(\rho) ,\mu}}=0\,.
\end{equation}

Any  transformation $\psi_{(\rho)}\to\psi'_{(\rho)}= \psi_{(\rho)}+\delta\psi_{(\rho)}$
leaving the action invariant,  ${\cal S}[\psi'_{(\rho)},\overline{\psi}'_{(\rho)}]={\cal
S}[\psi_{(\rho)},\overline{\psi}_{(\rho)}]$, is a symmetry transformation.
The Noether theorem shows that each symmetry transformation gives rise to the current
\begin{equation}
\Theta^{\mu}\,\propto\, \delta\overline{\psi}_{(\rho)}\,\frac{\partial {\cal
L}}{\partial \overline{\psi}_{(\rho),\mu}}+\frac{\partial {\cal L}}{\partial
{\psi}_{(\rho),\mu}}\,\delta\psi_{(\rho)}
\end{equation}
which is conserved in the sense that this satisfies $\Theta^{\mu}_{\,\,;\mu}=0$.

For the isometries transforming simultaneously the coordinates and the field components according to Eq. (\ref{es})  we have $\delta\psi_{(\rho)}=-i\xi^aX_a^{(\rho)}\psi_{(\rho)}$ where the operators $X_a^{(\rho)}$ are defined by Eq. (\ref{Xa}).
Consequently, each isometry of  parameter $\xi^a$ give rise to the corresponding conserved current
\begin{equation}
\Theta^{\mu}_a = i\left(\overline{X_a^{(\rho)}\psi_{(\rho)}}\,\frac{\partial {\cal
L}}{\partial {\overline{\psi}}_{(\rho),\mu}}-\frac{\partial {\cal L}}{\partial
{{\psi}}_{(\rho),\mu}}\, X_a^{(\rho)}\psi_{(\rho)}\right)\,, \quad a=1,2...N\,.
\end{equation}
Then we may define the relativistic scalar product $\langle~,~\rangle$ as
\begin{equation}
\langle\psi,\psi'\rangle=i \int_{\Sigma}
d\sigma_{\mu}\sqrt{g}\,\left(\overline{\psi}\,\frac{\partial {\cal
L'}}{\partial {\overline{\psi'}}_{,\mu}}-\frac{\partial {\cal
L}}{\partial {{\psi}}_{,\mu}}\,\psi'\right)\,,
\end{equation}
integrating on a time-like surface $\Sigma$ such that the conserved quantities (or charges) can be represented as expectation values, 
\begin{equation}\label{Ca}
C_a=\int_{\Sigma}
d\sigma_{\mu}\sqrt{g}\,\Theta_a^{\mu}=\langle\psi_{(\rho)},X_a^{(\rho)}\psi_{(\rho)}\rangle\,,
\end{equation}
of the isometry generators (\ref{Xa}).

We must specify that if the Lagrangian is invariant under isometries then the relativistic scalar product is also invariaant, $\langle T^{(\rho)}_{\frak g}\psi,T^{(\rho)}_{\frak g}\psi'\rangle=\langle\psi,\psi'\rangle$ while the operators (\ref{Xa})  are self-adjoint with respect to this scalar product,  $\langle X_a^{(\rho)}\psi,\psi'\rangle=\langle\psi,X_a^{(\rho)}\psi'\rangle$, since their spin parts are Dirac self-adjoint.

\subsection{Canonical quantization}

The operator Lie algebra generated by the isometry generators offers us the conserved operators that commutes with the operator of the field equation ${\cal E}$. Among them   we may select the sets of commuting operators $\{ A_1,A_2,...A_n\}$ determining the fundamental solutions (or quantum modes)  that satisfy the field equation and, in addition, the common eigenvalue problems
\begin{equation}\label{separ}
A_i U_{\alpha}=a_i U_{\alpha}\,, \quad A_i V_{\alpha}=- a_i V_{\alpha}\,,\quad i=1,2...n\,.
\end{equation}
corresponding to the eigenvalues $\alpha=\{a_1,a_2,...a_n\}$ from   the spectrum ${\Bbb S}={\Bbb S}_d\cup{\Bbb S}_c$ which may have a discrete part ${\Bbb S}_d$ and a continuous one ${\Bbb S}_c$.  We must stress that these 'dual' eigenvalues problems hold at any time when we use symmetric reps. $\rho$ allowing invariant forms. Then, according to Eq. (\ref{Cconj}) we have
\begin{equation}
(X_a^{(\rho)})^*=-C_{(\rho)}X_a^{(\rho)}C_{(\rho)}^{-1}\,,
\end{equation} 
 which means that the eigenvectors $U_{\alpha}$ and $V_{\alpha}$ are related through the {\em charge conjugation},  $V_{\alpha}=C_{(\rho)}U_{\alpha}^*$. For this reason, the solutions $U_{\alpha}$ of positive frequencies correspond to particles while the negative frequency ones describe antiparticles \cite{BDR}.

The set of solutions $U_{\alpha}$ form a basis of the particle subspace  ${\cal F}^+_{\rho}$  while the solutions $V_{\alpha}$ represent a basis of the corresponding antiparticle subspace ${\cal F}^-_{\rho}$. This separation is unique only when the commuting operators form a {\em complete} system. Otherwise, there are many possibilities of separation, each one defining its own vacuum state as it happens, for example, in the de Sitter case \cite{BD}. Then, supplemental criteria will help us to fix the vacuum state.

The fundamental solutions are orthogonal with respect to the relativistic scalar product and can be normalized such that
\begin{eqnarray}
&&\left<U_{\alpha},U_{\alpha'}\right>=\pm
\left<V_{\alpha},V_{\alpha'}\right>=
\delta(\alpha,\alpha')=\left\{
\begin{array}{ll}
\delta_{\alpha,\alpha'}\,,& \alpha,\alpha'\in {\Bbb S}_d\\
\delta(\alpha-\alpha')\,,& \alpha,\alpha'\in {\Bbb S}_c
\end{array}\right.
\label{orto1a}\\
&&\left<U_{\alpha},V_{\alpha'}\right>=
\left<V_{\alpha},U_{\alpha'}\right>=
0\,.\label{orto2a}
\end{eqnarray}
The spin-statistic connection requires to chose the sign $+$  for fermions and $-$ for bosons since then  the conserved quantities get a correct physical meaning in QFT.

The quantum fields can be expanded now in the above defined rep. $\alpha$ as  
\begin{equation}
\psi_{(\rho)}(x)=\psi_{(\rho)}^{(+)}(x)+\psi_{(\rho)}^{(-)}(x)=\int_{\alpha\in{\Bbb S}}U_{\alpha}(x)a(\alpha)+V_{\alpha}(x) b^{\dagger}(\alpha)\,,
\end{equation}
where we sum over the discrete part ${\Bbb S}_d$ and integrate over the continuous part ${\Bbb S}_c$ of the spectrum ${\Bbb S}$. The operators $a$ and $b$ are the particle and respectively antiparticle destruction (or annihilation) operators in rep. $\alpha$ \cite{BDR} 
for which we postulate  the canonical non-vanishing rules 
\begin{equation}\label{cac}
\left[a(\alpha),a^{\dagger}(\alpha')\right]_{\pm}=\left[b(\alpha),b^{\dagger}(\alpha')\right]_{\pm}=\delta(\alpha,\alpha')\,,
\end{equation}
where we denote $[x,y]_{\pm}=xy\pm yx$. In this manner,  the conserved quantities (\ref{Ca}) become  one-particle operators,
\begin{equation}\label{CXa}
C_a \to {\cal X}_a^{(\rho)} =:\langle\psi_{(\rho)},X_a^{(\rho)}\psi_{(\rho)}\rangle :\,,
\end{equation}
calculated respecting the normal ordering of the operator products \cite{BDR}. Thanks to this method of quantization,  the operators (\ref{CXa}) are now the isometry generators of the quantum field theory, forming the basis of the operator-valued rep.  of the $s(M)$ algebra. In a similar manner one can define the generators of the internal symmetries as, for example, the charge one-particle operator ${\cal Q}=:\langle\psi, \psi\rangle:$. 

Thus we obtain a rich operator algebra formed by field operators and one-particle operators which have the obvious properties
\begin{equation}\label{algXX}
\left[{\cal X}, \psi(x)\right]=-(X\psi)(x)\,, \quad
\left[{\cal X}, {\cal Y}\right]=:\left<\psi, ([X,Y]\psi)\right>: \,,
\end{equation} 
that preserve the linear structures. In general, for other algebraic relations we have a more complicated correspondence  between the algebra of the one-particle operators and that of the usual differential operators.  For example, the product of two one-particle operators of the form (\ref{CXa})  can be expanded as ${\cal X}{\cal Y}=:\langle \psi, XY\psi\rangle: +: {\cal X}{\cal Y}:$  where only the first term is an one-particle operator.  

Whether the one-particle operator $X$ does not mix among themselves  the subspaces of fundamental solutions ${\cal F}^+$ and ${\cal F}^-$ then it  can be expanded as
\begin{eqnarray}
{\cal X}&=&:\langle\psi,X\psi\rangle:={\cal X}^{(+)}+{\cal X}^{(-)}\nonumber\\
&=&
\int_{\alpha\in{\Bbb S}}\int_{\alpha'\in{\Bbb S}} \tilde X^{(+)}(\alpha,\alpha')a^{\dagger}(\alpha)a(\alpha')+ \tilde X^{(-)}(\alpha,\alpha')b^{\dagger}(\alpha)b(\alpha')\,,
\end{eqnarray}
where
\begin{equation}
\tilde X^{(+)}(\alpha,\alpha')=\langle U_{\alpha},X U_{\alpha'}\rangle\,, \quad
\tilde X^{(-)}(\alpha,\alpha')=\langle V_{\alpha},X V_{\alpha'}\rangle\,.
\end{equation} 
When the spectrum ${\Bbb S}={\Bbb S}_c$  is  continuous we may have  differential operators $\tilde X^{(\pm)}$ acting on the continuous variables  $\alpha$ such that
\begin{equation}
\tilde X^{(\pm)}(\alpha,\alpha')a^{\dagger}(\alpha)a(\alpha')=\delta(\alpha-\alpha')  a^{\dagger}(\alpha)\tilde X^{(\pm)}a(\alpha)\,.
\end{equation}
Then we say that  $\tilde X^{(\pm)}$ are the operators of the rep. $\alpha$ in the sense of the relativistic quantum mechanics. The best example is the momentum rep. largely used in the theory of the free fields on the Minkowski spacetime. Fortunalely, this rep. may be also used in the case of the de Sitter spacetime, as we shall see in what follows. 

The algebraic relations (\ref{cac}) remain invariant only if $a$ and $b$ transform according to UIRs of the isometry group.  A crucial problem is now the relation  between the CR transformimg the covariant field $\psi_{(\rho)}$ and the set of UIRs transforming the particle and antiparticle operators $a$ and $b$. This problem, referred here as the CR-UIR equivalence, is successfully solved in special relativity thanks to the Wigner theory of induced reps. of the Poincar\' e group.

\section{Covariant fields in special relativity}

On the Minkowski flat spacetime, $(M_0,\eta)$, the fields $\psi_{(\rho)}$ transform under isometries according to {\em manifest} covariant reps. in {\em inertial} (local) frames where the local indices coincide with the Cartesian natural ones such that  $e^{\mu}_{\nu}=\hat e^{\mu}_{\nu}=\delta^{\mu}_{\nu}$. The isometries  $x\to x'=\Lambda[A(\omega)]x
-a$ form the Poincar\' e group $I(M_0)= {\cal P}_{+}^{\uparrow} =
T(4)\,\circledS\, L_{+}^{\uparrow}$ \cite{W} whose universal covering
group is  $S(M_0)= \tilde{\cal P}^{\uparrow}_{+}=T(4)\,\circledS\,
SL(2,\Bbb C)$. Both these groups are semidirect products (denoted by  $\circledS$) where the translations form the {\em normal} Abelian subgroup $T(4)$.

\subsection{Generators of manifest CRs} 

The manifest CRs,  $ T^{(\rho)} \,:\,(A, a)\to  T_{A,a}^{(\rho)}$, of the group $S(M_0)$
have the transformation rules
\begin{equation}\label{TPoin}
(T^{(\rho)}_{A,a}\psi_{(\rho)})(x)
=\rho(A)\psi_{(\rho)}\left(\Lambda(A)^{-1}(x+a)\right)\,,
\end{equation}
and the well-known basis-generators of the $s(M_0)$ algebra,
\begin{eqnarray}
\hat P_{\mu}& \equiv &\hat X_{(\mu)}^{(\rho)}=i\partial_{\mu}\,, \label{XMin1}\\
\hat J_{\mu\nu}^{(\rho)}&\equiv&\hat X_{(\mu\nu)}^{(\rho)}
=i(\eta_{\mu\alpha}x^{\alpha}
\partial_{\nu}- \eta_{\nu\alpha}x^{\alpha}\partial_{\mu})+
S_{\mu\nu}^{(\rho)}\,,\label{XMin2}
\end{eqnarray}
which have point-independent spin parts $S_{\hat\mu\hat\nu}^{(\rho)}$. 
Here it is convenient to consider the standard momentum operator with contravariant components,  $\hat P^i=-i\partial_i$,  separating the energy operator, $\hat H=\hat
P_0=i\partial_t$, and writing the $sl(2,{\Bbb C})$ generators
\begin{eqnarray}
\hat J_i^{(\rho)}&=&\frac{1}{2}\,\varepsilon_{ijk}\hat
J^{(\rho)}_{jk}=
-i\varepsilon_{ijk}x^j\partial_k+S_i^{(\rho)}\,,\quad
S_i^{(\rho)}=\frac{1}{2}\,
\varepsilon_{ijk}S_{jk}^{(\rho)}\,,\\
\hat K_i^{(\rho)}&=&\hat J_{0i}^{(\rho)}=i (x^i
\partial_t+t\partial_i)+S_{0i}^{(\rho)}\,,\qquad i,j,k...=1,2,3\,,
\end{eqnarray}
that form the standard basis of the $s(M_0)$ algebra, $\{\hat H, \hat
P^i,\hat J^{(\rho)}_i,\hat K^{(\rho)}_i\}$ \cite{WKT}.

The invariants of the manifest covariant fields are  the eigenvalues
of the Casimir operators of the reps. $T^{(\rho)}$ that
read
\begin{equation}\label{CM1}
\hat {C}_1=\hat P_{\mu}\hat P^{\mu}\,,\quad \hat {
C}_2^{(\rho)}=-\eta_{\mu\nu}\hat W^{(\rho)\,\mu}\hat
W^{(\rho)\,\nu}\,,
\end{equation}
where the Pauli-Lubanski operator \cite{W},
\begin{equation}\label{PaLu}
\hat
W^{(\rho)\,\mu}=-\frac{1}{2}\,\varepsilon^{\mu\nu\alpha\beta}\hat
P_{\nu} \hat J_{\alpha\beta}^{(\rho)}\,,
\end{equation}
has the components
\begin{equation}
\hat W^{(\rho)}_0={\hat J}^{(\rho)}_i{\hat P}_i={S}^{(\rho)}_i{\hat
P}_i\,,\quad \hat W^{(\rho)}_i=\hat H\,{\hat J}_i^{(\rho)}+
\varepsilon_{ijk}{\hat K}^{(\rho)}_j {\hat P}_k\,,
\end{equation}
resulting from Eqs. (\ref{XMin1}) and (\ref{XMin2}) where we
take $\varepsilon^{0123}=-\varepsilon_{0123}=-1$.

In the Poincar\' e algebra we find the complete system of commuting operators $\{\hat H, \hat P^1, \hat P^2,\hat P^3\}$ defining the momentum rep.. The fundamental solutions are common eigenfunctions of  this system, corresponding to the eigenvalues $\{E, p^1, p^2,p^3\}$ such that any covariant quantum field can be written as
\begin{equation}\label{Cfield}
\psi_{(\rho)}(x)=\int d^3p \sum_{s\sigma}\left[U_{{\bf p},s\sigma}(x) a_{s \sigma}({\bf p}) +V_{{\bf p},s\sigma}(x) b^{\dagger}_{s \sigma}({\bf p})\right]
\end{equation}
where $a_{s\sigma}$ and $b_{s\sigma}$ are the field operators of a particle and antiparticle of {\em spin} $s$ and {\em polarization} $\sigma$ while the fundamental solutions have the form
\begin{equation}\label{Fsol}
U_{{\bf p},s\sigma}(x)=\frac{1}{(2\pi)^{\frac{3}{2}}}u_{s\sigma}({\bf p}) e^{-iEt+i{\bf p}\cdot{\bf x}}\,, \quad
V_{{\bf p},s\sigma}(x)=\frac{1}{(2\pi)^{\frac{3}{2}}}v_{s\sigma}({\bf p}) e^{iEt-i{\bf p}\cdot{\bf x}}\,.
\end{equation} 
The vectors $u_{s\sigma}({\bf p})$ and $v_{s\sigma}({\bf p})$ have to be determined by the concrete forms of the field equation and relativistic scalar product. However, when the field equations are linear we can postulate orthonormalization relations of the general form  
\begin{eqnarray}
&&\overline{u}_{s\sigma}({\bf p}) u_{s'\sigma'}({\bf p}) =\pm\overline{v}_{s\sigma}({\bf p}) v_{s'\sigma'}({\bf p})=N({\bf p})^2\delta_{ss'}\delta_{\sigma\sigma'}\,,\label{uuvv}\\
&&\overline{u}_{s\sigma}({\bf p}) v_{s'\sigma'}({\bf p}) =\overline{v}_{s\sigma}({\bf p}) u_{s'\sigma'}({\bf p})=0\,.\label{uvuv}
\end{eqnarray}
where $N({\bf p})$ is a normalization factor that satisfies $N(0)=1$ and may depend on the representation $\rho$ and the form of relativistic scalar product. These relations guarantee the separation of the particle and antiparticle sectors of any covariant field.

The first invariant (\ref{CM1}a) gives the mass condition, $\hat
P^2\psi_{(\rho)}=m^2\psi_{(\rho)}$, fixing the orbit in the momentum
spaces on which the fundamental solutions are defined.  For the massive fields of mass $m$ the momentum spans the orbit  $\Omega_m=\{{\bf p}\,|\, p^2=m^2\}$ which means that $p_0=\pm E$ where $E=E({\bf p})=\sqrt{m^2 +{\bf p}^2}$.  The solutions $U$ are considered of positive frequencies having $p_0=E$ while for the negative frequency ones, $V$, we must take $p_0=-E$. In this manner the general rule (\ref{separ}) of separating the particle and antiparticle modes becomes
\begin{eqnarray}
&&\hat H U_{{\bf p},s\sigma}=E U_{{\bf p},s\sigma}\,,\quad 
\hat H V_{{\bf p},s\sigma}=-E V_{{\bf p},s\sigma}\,,\\
&&{\hat P}^i U_{{\bf p},s\sigma}={p}^i\, U_{{\bf p},s\sigma}\,,\quad ~
{\hat P}^i V_{{\bf p},s\sigma}=-{p}^i\, V_{{\bf p},s\sigma}\,.
\end{eqnarray}

The second invariant is less relevant for the CRs since its form in configuration rep. is quite complicated as long as it reads
\begin{eqnarray}
\hat C_2^{(\rho)}&=& -({\bf S}^{(\rho)})^2\partial_t^2 +
2 (iS_{0k}^{(\rho)}-\varepsilon_{ijk}S_i^{(\rho)}
S_{0j}^{(\rho)})\partial_k\partial_t\nonumber\\
&&-\left[({\bf S_0}^{(\rho)})^2\Delta
-(S_i^{(\rho)}S_j^{(\rho)}+S_{0i}^{(\rho)}S_{0j}^{(\rho)})
\partial_i\partial_j\right]\,,\label{Q220}
\end{eqnarray}
where we denote  ${\bf S}^2=S_iS_i$ and ${\bf S}_0^2=S_{0i}S_{0i}$. 
Consequently, we may study its action in  momentum rep. where it selects the induced Wigner UIRs equivalent with the manifest CRs.

\subsection{Wigner's induced UIRs}

The Wigner theory of  the induced UIRs is based on the fact that an orbit in momentum space may be built by using Lorentz transformations \cite{Wigner, Mck}.  In the case of massive particles we discuss here,  any ${\bf p}\in \Omega_m$ can be obtained applying a suitable {\em boost} transformation $L_{{\bf p}}\in L_{+}^{\uparrow}$ to the {\em representative} momentum $\mathring{p}=(m,0,0,0)$ such that  ${\bf p}=L_{{\bf p}}\,\mathring{p}$. The rotations that leave $\mathring{p}$ invariant, $R\mathring{p}=\mathring{p}$, form the {\em stable} group $SO(3)\subset L_{+}^{\uparrow}$ whose  universal covering group $SU(2)$  is called the {\em little} group associated to the representative momentum $\mathring{p}$. 

We observe that  the boosts $L_{{\bf p}}$ are defined up to a rotation since $L_{{\bf p}}R\,\mathring{p}=L_{{\bf p}}\,\mathring{p}$ such that  these form the homogeneous space $L_{+}^{\uparrow}/SO(3)$.  The corresponding transformations of the $SL(2,{\Bbb C})$ group are denoted by $A_{{\bf p}}\in SL(2,{\Bbb C})/SU(2)$ assuming that these satisfy 
$\Lambda(A_{{\bf p}})=L_{{\bf p}}$ and $A_{\mathring{p}}=1\in SL(2,{\Bbb C})$. 

In applications one prefers to chose genuine Lorentz transformatios  \cite{Th}
\begin{equation}\label{roAp}
A_{{\bf p}}=\exp\left(-i  n^i S_{0i}\,{\rm arctanh} \frac{p}{E}\right)=\frac{E+m+\gamma^0\gamma^i p^i}{\sqrt{2m(E+m)}}\,,
\end{equation}
where  $n^i=\frac{p^i}{p}$ with the notation $p=|{\bf p}|$.  The corresponding transformations of the $L_{+}^{\uparrow}$ group, $L_{{\bf p}}=\Lambda(A_{{\bf p}})$, have the matrix elements 
\begin{equation}
(L_{{\bf p}})^{0\,\cdot}_{\cdot\, 0}=\frac{E}{m}\,,\quad (L_{{\bf p}})^{0\,\cdot}_{\cdot\, i}=(L_{{\bf p}})^{i\,\cdot}_{\cdot\, 0}=\frac{p^i}{m}\,,\quad 
(L_{{\bf p}})^{i\,\cdot}_{\cdot\, j}=\delta_{ij}+\frac{p^i p^j}{m(E+m)}\,.
\end{equation}

Furthermore, we look for  the transformations in momentum rep. generated by the transformation rule (\ref{TPoin}) of the manifest CR under consideration.  After a little calculation we obtain
\begin{eqnarray}
&&\sum_{s' \sigma'}u_{s' \sigma'}({\bf p})(T_{A,a} a_{s' \sigma'})({\bf p})
=\frac{E'}{E}\sum_{s \sigma} \rho(A)u_{s \sigma}\left( {{\bf p}\,}'\right)a_{s\sigma}\left({{\bf p}\,}'\right) e^{-i {a}\cdot{p}}\\
&&\sum_{s' \sigma'}v_{s' \sigma'}({\bf p})(T_{A,a} b^{\dagger}_{s' \sigma'})({\bf p})
=\frac{E'}{E}\sum_{s \sigma} \rho(A)v_{s \sigma}\left({{\bf p}\,}'\right)b^{\dagger}_{s\sigma}\left( {{\bf p}\,}'\right) e^{i {a}\cdot{p}}
\end{eqnarray}
where $a\cdot p=Ea^0-{\bf p}\cdot {\bf a}$ and ${{\bf p}\,}'=\Lambda(A)^{-1}{\bf p}~\to~ E'=E({\bf p}')$. Focusing on the first equation,  we introduce the Wigner particle mode vectors, 
\begin{equation}\label{WigU}
u_{s\sigma}({\bf p})=N({\bf p}) \rho(A_{{\bf p}}) \mathring{u}_{s\sigma}\in{\cal V}_{(\rho)}\,,
\end{equation}
where the vectors $\mathring{u}_{s\sigma}\in {\cal V}_{(\rho)}$ are independent on ${\bf p}$ and satisfy 
${\mathring{u}}^+_{s\sigma}\mathring{u}_{s'\sigma'}=\delta_{ss'}\delta_{\sigma\sigma'}$ while $N({\bf p})=N(p)$ is a normalization factor depending only on $p=|{\bf p}|$.   We obtain thus the transformation rules  of the Wigner irreps. of spin $s$ {\em  induced} by the subgroup $T(4)\,\circledS\,SU(2)$, that read \cite{Wigner,WKT}
\begin{equation}\label{Wig}
(T_{A,a} a_{s\sigma})({\bf p})=\frac{E({p}')N({{p}'})}{E({p})N({p})}\sum_{\sigma'} D^s_{\sigma\sigma'}(A,{\bf p}) a_{s\sigma'}({{\bf p}\,}') e^{i{\bf a}\cdot{\bf p}}
\end{equation}
where
\begin{equation}
D^s_{\sigma\sigma'}(A,{\bf p})={\mathring{u}}^+_{s\sigma}\rho[W(A,{\bf p})]\mathring{u}_{s\sigma'}\,, \quad W(A,{\bf p})=A^{-1}_{{\bf p}} A A_{{{\bf p}\,}'}
\end{equation} 
The Wigner transformations $W(A,{\bf p})=A^{-1}_{{\bf p}} A A_{{{\bf p}\,}'}$ are transformations of the little group $SU(2)$. Indeed, one can verify that the transformations $\Lambda[W(A,{\bf p})]=L^{-1}_{{\bf p}}\Lambda(A) L_{{{\bf p}\,}'}$ leave invariant the representative momentum, 
\begin{equation}
L^{-1}_{{\bf p}}\Lambda(A) L_{{{\bf p}\,}'}\mathring{p}=L^{-1}_{{\bf p}}\Lambda(A){{\bf p}\,}' =L^{-1}_{{\bf p}}{\bf p}=\mathring{p}\,,
\end{equation}
which means that $\Lambda[W(A,{\bf p})]\in SO(3) \to W(A,{\bf p})\in SU(2)$.

The conclusion is that the matrices $D^s$ realize the UIRs of spin $s$ of the little group $SU(2)$ that induces the Wigner UIR  (\ref{Wig}) \cite{WKT}. The role of the vectors $\mathring{u}_{s\sigma}$ is to select the spin content of the CR determining the Wigner UIRs whose direct sum is equivalent to the manifest CR $T^{(\rho)}$. 

A similar procedure can be applied for the antiparticle defining the corresponding vectors
\begin{equation}\label{WigV}
v_{s\sigma}({\bf p})=N({\bf p}) \rho(A_{{\bf p}}) \mathring{v}_{s\sigma}\in{\cal V}_{(\rho)}\,,
\end{equation}
but selecting the normalized vectors $\mathring{v}_{s\sigma}\in{\cal V}_{(\rho)}$ such that  ${\mathring{v}}^+_{s\sigma}\mathring{v}_{s'\sigma'}=\delta_{ss'}\delta_{\sigma\sigma'}$ and \cite{W}
\begin{equation}
{\mathring{v}}^+_{s\sigma}\rho[W(A,{\bf p})]\mathring{v}_{s\sigma'}= [D^s_{\sigma\sigma'}(A,{\bf p})]^*  
\end{equation}
since the operators $a$ and $b$ must transform alike under isometries \cite{W}.  Moreover,  from Eq. (\ref{uvuv}) we deduce that the vectors $\mathring{u}$ and $\mathring{v}$ must be orthogonal, ${\mathring{u}}^+_{s\sigma}\mathring{v}_{s'\sigma'}={\mathring{v}}^+_{s\sigma}\mathring{u}_{s'\sigma'}=0$. We remind the reader that the $SU(2)$ UIRs have the equivalence property \cite{WKT}
\begin{equation}\label{DY}
(D^s)^*=(Y^s)^{-1} D^s Y^s\,, \quad Y^s_{\sigma,\sigma'}=(-1)^{s-\sigma}\delta_{\sigma,-\sigma'}\,.
\end{equation}

Thus we demonstrated  that the manifest CRs are equivalent to direct sums of Wigner UIRs with an arbitrary spin content defined by the vectors $ \mathring{u}_{s\sigma}$ and $\mathring{v}_{s\sigma}$. For each spin $s$ we meet the induced UIR carried by the space ${\cal V}_s\subset {\cal V}_{(\rho)}$ of the linear UIR of the group $SU(2)$ generated by the matrices $S_i^{(s)}$.  

The transformation (\ref{Wig}) allows us to derive the generators of the UIRs in momentum rep. (denoted here by tilde) that are differential operator acting alike on the operators $a_{s\sigma}({\bf p})$ and $b_{s\sigma}({\bf p})$ seen as functions of ${\bf p}$. Thus for each UIR $(\pm m,s)$ we can write down the basis generators 
\begin{eqnarray}
\tilde J_i^{(s)}&=&-i \varepsilon_{ijk}p^j\partial_{p^k}+S_i^{(s)}\,,\\
\tilde K_i^{(s)}&=&iE \partial_{p^i} +\frac{1}{N}\partial_{p^i}(EN)+\frac{1}{E+m}\,\varepsilon_{ijk}p^jS_k^{(s)}\,.
\end{eqnarray}
With their help we derive the components of the Pauli-Lubanski operator
\begin{equation}
\tilde W_0^{(s)}={\bf p}\cdot {\bf S}^{(s)}\,, \quad \tilde W_i^{(s)}=m S^{(s)}_i+\frac{p^i}{E+m}{\bf p}\cdot {\bf S}^{(s)}\,,
\end{equation}
and we recover the well-known result $\tilde {C}_2^{(s)}= m^2 ({\bf S}^{(s)})^2\sim m^2 s(s+1)$ \cite{WKT}.

Finally we stress that the Wigner theory determine completely the form of the covariant fields without using field equations.  Thus in special relativity we have two symmetric equivalent procedures: (i) to start with the covariant field equation that gives the form of the covariant field determining thus its manifest CR, or (ii) to construct the Wigner covariant field and then to derive its field equation \cite{W}. 

\subsection{Example: the Dirac field}

The free Dirac field on the Minkowski spacetime is a typical example of manifest covariant free field with unique spin $s=\frac{1}{2}$ that can be constructed following the method (ii) \cite{WKT,Th}. In this case the CR is induced by the rep. $\rho_D= =(\frac{1}{2},0)\oplus(0,\frac{1}{2})$ which, as mentioned before,  is considered here as the fundamental rep. of the $SL(2,{\Bbb  c})$ group generated by the matrices (\ref{Sgg}).

According to Eqs. (\ref{WigU}) and (\ref{WigV}), the Dirac spinors $u_{\sigma}({\bf p})$ and $v_{\sigma}({\bf p})$ can be written as 
\begin{equation}\label{muc}
u_{\sigma}({\bf p})=N({\bf p}) A_{{\bf p}} \mathring{u}_{s\sigma}\,,\quad 
v_{\sigma}({\bf p})=N({\bf p}) A_{{\bf p}} \mathring{v}_{s\sigma}\,,
\end{equation}
where the matrix $A_{{\bf p}}$ is defined by Eq. (\ref{roAp}) while the constant spinors 
\begin{equation}
\mathring{u}_{\sigma}=\frac{1}{\sqrt{2}}\left(
\begin{array}{c}
\xi_{\sigma}\\
\xi_{\sigma}
\end{array}\right)\,,\quad
\mathring{v}_{\sigma}=\frac{1}{\sqrt{2}}\left(
\begin{array}{c}
\eta_{\sigma}\\
-\eta_{\sigma}
\end{array}\right)\,,
\end{equation}
depend on the Pauli spinors $\xi_{\sigma}$ and  $\eta_{\sigma}=Y\xi_{\sigma}^*=i\sigma_2 \xi_{\sigma}^*$ which satisfy the condition  (\ref{DY}). Thus we understand that  only the first Pauli spinor, $\xi_{\sigma}$, can be chosen arbitrarily, determining in this manner the type of basis. 

In practice one use two such bases of the two-dimensional spinor spaces carrying the fundamental repr. of the $SU(2)$ group generated by the matrices $S_i^{(\frac{1}{2})}=\frac{1}{2}\sigma_i$. The first one,  called the {\em spin basis}, is formed by the eigenspinors of the matrix $S^{(\frac{1}{2})}_3$ which satisfy  $S^{(\frac{1}{2})}_3\xi_{\sigma}=\sigma \xi_{\sigma}$ having the form
\begin{equation}
\xi_{\frac{1}{2}}=\left(
\begin{array}{c}
1\\
0
\end{array}\right)\,,\quad
\xi_{-\frac{1}{2}}=\left(
\begin{array}{c}
0\\
1
\end{array}\right)\,.
\end{equation}  
Another basis is the {\em helicity} one whose spinors  fulfill 
\begin{equation}
{{\bf S}}^{(\frac{1}{2})}\cdot {\bf p}\,\xi_{\lambda}({\bf p})=|{\bf p}| \lambda\, \xi_{\lambda}({\bf p})\,,
\end{equation}
where $\lambda=\pm \frac{1}{2}$ is the helicity. 

With these preparations the Dirac field can be written in the form (\ref{Cfield}) as 
\begin{equation}\label{CfieldD}
\psi_D(x)=\int d^3p \sum_{\sigma}\left[U_{{\bf p},\sigma}(x) a_{ \sigma}({\bf p}) +V_{{\bf p},\sigma}(x) b^{\dagger}_{\sigma}({\bf p})\right]
\end{equation}
where $a_{\sigma}$ and $b_{\sigma}$ are the field operators of a particle and antiparticle of {spin} $s=\frac{1}{2}$ and {polarization} $\sigma=\pm\frac{1}{2}$ while the fundamental solutions 
\begin{equation}\label{FsolD}
U_{{\bf p},\sigma}(x)=\frac{1}{(2\pi)^{\frac{3}{2}}}u_{\sigma}({\bf p}) e^{-iEt+i{\bf p}\cdot{\bf x}}\,, \quad
V_{{\bf p},\sigma}(x)=\frac{1}{(2\pi)^{\frac{3}{2}}}v_{\sigma}({\bf p}) e^{iEt-i{\bf p}\cdot{\bf x}}\,,
\end{equation} 
are completely determined up to the normalization factor $N({\bf p})$ which remains arbitrary. 

The next step is to look for the field equation in momentum representation and then to establish the final form of the field equation in configuration. 
For this purpose we start with the observation that the matrices (\ref{roAp}) and $\gamma p=E\gamma^0-{\bf \gamma}\cdot{\bf p}$ satisfy  the identity
$\gamma p\,A_{{\bf p}}=mA_{{\bf p}}\gamma^0$ which allows us to write the equations
\begin{equation}
(\gamma p-m)u_{\sigma}({\bf p})=0\,, \quad (\gamma p+m)v_{\sigma}({\bf p})=0\,,
\end{equation}
 since $\gamma^0\mathring{u}_{\sigma}=\mathring{u}_{\sigma}$ and $\gamma^0\mathring{v}_{\sigma}=-\mathring{v}_{\sigma}$. The conclusion is that the spinors (\ref{FsolD}) are the fundamental solutions of the Dirac equation
 \begin{equation}
 \left(i\gamma^{\mu}\partial_{\mu}-m\right)\psi_D=0\,.
 \end{equation}
Thus the Dirac equation can be obtained by using exclusively group theoretical methods.
However, these are not enough for a field theory since we need, in addition, to have a relativistic scalar product derived from a Lagrangian theory.  The orthonormalization condition with respect to this scalar product has to determine the factor $N({\bf p})$ of Eq. (\ref{muc}) \cite{WKT}.

\section{Covariant fields on de Sitter spacetime}

The Wigner theory works only in  local-Minkowskian manifold whose isometry group has a similar structure as the Poincar\' e one, having an Abelian normal (or invariant) subgroup $T(4)$.   Unfortunately,  the Abelian group $T(3)_P$ of the de Sitter isometry group $SO(1,4)$  is not a normal subgroup such that the Wigner method does not work on this manifold. 

An alternative method was proposed by Nachtmann that constructed the CRs on the de Sitter spacetime observing that this is isomorphic with the coset space $SO(1,4)/L_{+}^{\uparrow}$ that form an orbit in $M^5$ where the Wigner method can be adapted but in configuration instead of momentum rep. We have shown  \cite{CNach} that these CRs are just the general ones which will be presented in what follows  in the configuration rep. We shall see that the theory the UIRs in momentum rep. and, implicitly, the CR-UIR equivalence may be considered only when the structure of the covariant field is determined by a concrete field equation. 

\subsection{de Sitter isometries and Killing vectors}

Let us start with the de Sitter spacetime $(M,g)$ defined as the hyperboloid of radius $1/\omega$ \footnote{We denote by $\omega$ the Hubble de Sitter constant since  $H$ is reserved for the energy operator} in the five-dimensional flat spacetime $(M^5,\eta^5)$ of coordinates $z^A$  (labeled by the indices $A,\,B,...= 0,1,2,3,4$) and metric $\eta^5={\rm diag}(1,-1,-1,-1,-1)$. The local charts $\{x\}$  can be introduced on $(M,g)$ giving the set of functions $z^A(x)$ which solve the hyperboloid equation,
\begin{equation}\label{hip}
\eta^5_{AB}z^A(x) z^B(x)=-\frac{1}{\omega^2}\,.
\end{equation}
Here we use the chart $\{t,{\bf x}\}$ with the conformal time $t$ and Cartesian spaces coordinates $x^i$ defined by
\begin{eqnarray}
z^0(x)&=&-\frac{1}{2\omega^2 t}\left[1-\omega^2({t}^2 - {\bf x}^2)\right]
\nonumber\\
z^i(x)&=&-\frac{1}{\omega t}x^i \,, \label{Zx}\\
z^4(x)&=&-\frac{1}{2\omega^2 t}\left[1+\omega^2({t}^2 - {\bf x}^2)\right]
\nonumber
\end{eqnarray}
This chart  covers the expanding part of $M$ for $t \in (-\infty,0)$
and ${\bf x}\in {\Bbb R}^3$ while the collapsing part is covered by
a similar chart with $t >0$. Both these charts have the
conformal flat line element,
\begin{equation}\label{mconf}
ds^{2}=\eta^5_{AB}dz^A(x)dz^B(x)=\frac{1}{\omega^2 {t}^2}\left({dt}^{2}-d{\bf x}^2\right)\,.
\end{equation}
In addition, we consider the local frames $\{t,{\bf x};e\}$ of the diagonal gauge,
\begin{equation}\label{tt}
e^{0}_{0}=-\omega t\,, \quad e^{i}_{j}=-\delta^{i}_{j}\,\omega t
\,,\quad
\hat e^{0}_{0}=-\frac{1}{\omega t}\,, \quad \hat e^{i}_{j}=-\delta^{i}_{j}\,
\frac{1}{\omega t}\,.
\end{equation} 

The gauge group $G(\eta^5)=SO(1,4)$  is the isometry group of $M$,  since its transformations, $z\to {\frak g}z$,  ${\frak g}\in SO(1,4)$, leave the Eq. (\ref{hip}) invariant. Its universal covering group ${\rm Spin}(1,4)=Sp(2,2)$ is not involved directly in our construction since the spinor CRs are induced by the spinor reps. of its subgroup $SL(2,\Bbb C)$. Therefore, we can restrict ourselves to the group $SO(1,4)$ for which we adopt the parametrization
\begin{equation}
{\frak g}(\xi)=\exp\left(-\frac{i}{2}\,\xi^{AB}{\frak S}_{AB}\right)\in SO(1,4) 
\end{equation}
with skew-symmetric parameters, $\xi^{AB}=-\xi^{BA}$,  and the covariant generators ${\frak S}_{AB}$ of the fundamental rep. of the $so(1,4)$ algebra carried by $M^5$. These generators have the matrix elements, 
\begin{equation}
({\frak S}_{AB})^{C\,\cdot}_{\cdot\,D}=i\left(\delta^C_A\, \eta_{BD}^5
-\delta^C_B\, \eta_{AD}^5\right)\,.
\end{equation}
The principal $so(1,4)$ basis-generators with physical meaning \cite{CCC} are the energy ${\frak H}=\omega{\frak S}_{04}$, angular momentum  ${\frak J}_k=\frac{1}{2}\varepsilon_{kij}{\frak S}_{ij}$, Lorentz boosts ${\frak K}_i={\frak S}_{0i}$, and the Runge-Lenz-type vector ${\frak R}_i={\frak S}_{i4}$. In addition, it is convenient to introduce the momentum ${\frak P}_i=-\omega({\frak R}_i+{\frak K}_i)$ and its dual ${\frak Q}_i=\omega({\frak R}_i-{\frak K}_i)$ which are nilpotent 
matrices (i. e. $({\frak P}_i)^3=({\frak Q}_i)^3=0$) of two Abelian three-dimensional subalgebras, $t(3)_P$ and respectively $t(3)_Q$ generating the Abelian subgroups $T(3)_P$ and $T(3)_Q$.  Among all these generators we may chose different bases of the algebra $so(1,4)$ as, for example, the basis $\{{\frak H},{\frak P}_i,{\frak Q}_i,{\frak J}_i\}$ or the Poincar\' e-type one, $\{{\frak H},{\frak P}_i,{\frak J}_i,{\frak K}_i\}$. We note that the four-dimensional restriction of the $so(1,3)$ subalegra generate the vector rep. of the group $L_+^{\uparrow}$.

Using these generators we can derive the $SO(1,4)$ isometries, $x\to x'=\phi_{\frak g}(x)$, that can be obtained solving the system   
\begin{equation}\label{zgz}
z[\phi_{\frak g}(x)]={\frak g}\,z(x). 
\end{equation}
The transformations ${\frak g}\in SO(3)\subset SO(4,1)$  generated by ${\frak J}_i$, are simple rotations of $z^i$ and $x^i$ which transform alike since this symmetry is global. The transformations generated by ${\frak H}$,
\begin{equation}\label{transH}
\exp(-i\xi {\frak H})\,:\quad
\begin{array}{lcl}
z^0&\to&z^0 \cosh\alpha-z^4 \sinh\alpha \\
z^i&\to&z^i\\
z^4&\to&-z^0 \sinh\alpha+z^4 \cosh\alpha 
\end{array}
\end{equation}
whith $\alpha=\omega\xi$, produce the dilatations
$t\to t\,e^{\alpha}$ and $x^i\to x^i e^{\alpha}$,
while the $T(3)_P$ transformations
\begin{equation}\label{transP}
\exp(-i{\xi}^i{\frak P}_i)\,:\quad
\begin{array}{lcl}
z^0&\to&z^0 +\omega\,{\bf \xi}\cdot{\bf z}+\frac{1}{2}\,\omega^2
{{\bf \xi}\,}^2\,(z^0+z^4) \\
z^i&\to&z^i+\omega\,\xi^i\,(z^0+z^4)\\
z^4&\to&z^4 -\omega\,{\bf \xi}\cdot{\bf z}-\frac{1}{2}\,\omega^2
{{\bf \xi}\,}^2\,(z^0+z^4) 
\end{array}
\end{equation}
give rise to the space translations $ x^i\to x^i +\xi^i$ at fixed $t$.
More interesting are the $T(3)_Q$ transformations generated by ${\frak Q}_i/\omega$,
\begin{equation}\label{transN}
\exp(-i{\xi}^i {\frak Q}_i/\omega)\,:\quad
\begin{array}{lcl}
z^0&\to&z^0 -{\bf \xi}\cdot{\bf z}+\frac{1}{2}\,
{{\bf \xi}\,}^2\,(z^0-z^4) \\
z^i&\to&z^i-\xi^i\,(z^0-z^4)\\
z^4&\to&z^4 -{\bf \xi}\cdot{\bf z}+\frac{1}{2}\,
{{\bf \xi}\,}^2\,(z^0-z^4) 
\end{array}
\end{equation}
which lead to the isometries
\begin{eqnarray}
t&\to&\frac{t}{1-2\omega\, {\bf \xi}\cdot{\bf x}
-\omega^2{{\bf \xi}\,}^2\,({t}^2-{\bf x}^2)} \\
x^i&\to&\frac{x^i+\omega\xi^i\, ({t}^2-{\bf x}^2)}
{1-2\omega\, {\bf \xi}\cdot{\bf x}
-\omega^2{{\bf \xi}\,}^2\,({t}^2-{\bf x}^2)}\,.
\end{eqnarray}
We observe that $z^0+z^4=-\frac{1}{\omega^2 t}$ is invariant under translations (\ref{transP}), fixing the value of $t$, while $z^0-z^4=\frac{t^2-{\bf x}^2}{t}$ is left unchanged by the $t(3)_Q$ transformations (\ref{transN}).

The orbital basis-generators of the natural rep. of the $s(M)$ algebra (carried by the space of the scalar functions over $M^{5}$) have the standard form
\begin{equation}\label{LAB5}
 L_{AB}^{5}=i\left[\eta_{AC}^5 z^{C}\partial_{B}-
 \eta_{BC}^5 z^{C}
\partial_{A}\right]=-iK_{(AB)}^C\partial_C
\end{equation}
which allows us to derive the corresponding Killing vectors of
$(M,g)$, $k_{(AB)}$, using the identities $k_{(AB)
\mu}dx^{\mu}=K_{(AB) C}dz^C$. Thus we obtain the following components of the Killing vectors:
\begin{eqnarray}
&&k_{(04)}^0=t\,, \quad k_{(04)}^i=x^i\,,\quad  k_{(0i)}^0=k_{(4i)}^0=\omega t x^i\,,\label{chichi1}\\
&&k_{(0i)}^j=\omega x^ix^j+\delta_i^j\frac{1}{2\omega}[\omega^2(t^2-{{\bf x}}^2)-1]\,,\\
&&k_{(4i)}^j=\omega x^ix^j+\delta_i^j\frac{1}{2\omega}[\omega^2(t^2-{{\bf x}}^2)+1]\,,\\
&&k_{(ij)}^k=\delta^k_jx^i-\delta^k_i x^j\,,\label{chichi2}
\end{eqnarray}
which will help us to derive the conserved observables.

\subsection{Generators of induced CRs}

The covariant field transform under the isometries defined by Eq. (\ref{zgz}) according to the general rule (\ref{Tx}). Therefore, in the covariant parametrization of the $sp(2,2)$ algebra, adopted here, the generators $X_{(AB)}^{(\rho)}$ corresponding to the
Killing vectors $k_{(AB)}$  result from  Eq. (\ref{Xa}) and the functions (\ref{Om}) with the new labels $a\to (AB)$. Using then the components (\ref{chichi1}) - (\ref{chichi2}) and the tetrad-gauge (\ref{tt}) of the chart $\{t,{\bf x}\}$, after a little calculation, we find the concrete form of the $sl(2,{\Bbb C})$ generators of the CRs induced by the reps. $\rho$.   
These are differential matrix-operators, satisfying the canonical commutation rules of the $so(1,4)$ algebra \cite{WKT}, that may be interpreted as  energy (or Hamiltonian) $H$,  total angular momentum ${\bf J}$, generators of the Lorentz boosts ${\bf K}$, and a Runge-Lenz type vector ${\bf R}$, whose components read \cite{CCC},
\begin{eqnarray}
H& =&  \omega X_{(04)}^{(\rho)}=-i\omega(t \partial_t + {x}^i
{\partial}_i)\,,\label{Ham}\\
J_i ^{(\rho)}&=&  \frac{1}{2}\,\varepsilon_{ijk}X_{(jk)}^{(\rho)}=-i\varepsilon_{ijk}x^j\partial_k+S_i^{(\rho)}\,,\label{Ji}\\
K_i^{(\rho)}& =&  X_{(0i)}^{(\rho)}=i x^i H+
\frac{i}{2\omega}[1+\omega^2( {\bf x}^2-t^2)]\partial_i -\omega
tS_{0i}^{(\rho)}+\omega S_{ij}^{(\rho)}x^j\,,\label{Ki}\\
R_i^{(\rho)}  &=&X_{(i4)}^{(\rho)}=-K_i^{(\rho)}+\frac{1}{\omega}\, i\partial_i\,.\label{Ri}
\end{eqnarray}
These generators form the basis $\{
H,J^{(\rho)}_i,K^{(\rho)}_i,R^{(\rho)}_i\}$ of the induced CR of the $sp(2,2)$ algebra with the following commutation rules:
\begin{eqnarray} &\left[ J_i^{(\rho)},
J_j^{(\rho)}\right]=i\varepsilon_{ijk} J_k^{(\rho)}\,,\quad&\left[
J_i^{(\rho)},
R_j^{(\rho)}\right]=i\varepsilon_{ijk} R_k^{(\rho)}\,,\label{ALG1}\\
&\left[ J_i^{(\rho)}, K_j^{(\rho)}\right]=i\varepsilon_{ijk}
K_k^{(\rho)}\,,\quad&\left[ R_i^{(\rho)},
R_j^{(\rho)}\right]=i\varepsilon_{ijk}
J_k^{(\rho)}\,,\label{ALG2}\\
&\left[ K_i^{(\rho)}, K_j^{(\rho)}\right]=-i\varepsilon_{ijk}
J_k^{(\rho)}\,,\quad&\left[ R_i^{(\rho)},
K_j^{(\rho)}\right]=\frac{i}{\omega}\,\delta_{ij}H\,,\label{ALG3}
\end{eqnarray}
and
\begin{equation}\label{HHKR}
\left[ H, J_i^{(\rho)}\right]=0\,,\quad \left[ H,
K_i^{(\rho)}\right]=i\omega R_i^{(\rho)}\,,\quad\left[ H,
R_i^{(\rho)}\right]=i\omega K_i^{(\rho)}\,.
\end{equation}

In some applications it is useful to replace the operators
${\bf K}^{(\rho)}$ and ${\bf R}^{(\rho)}$ by the Abelian ones, i. e. the momentum operator ${\bf P}$ and its dual ${\bf Q}^{(\rho)}$, whose covariant components are
defined as
\begin{equation}\label{Pi}
P_i= \omega(R^{(\rho)}_i+K^{(\rho)}_i)=i\partial_i\,, \quad
Q^{(\rho)}_i= \omega(R^{(\rho)}_i-K^{(\rho)}_i)\,.
\end{equation}
Bering in mind that here the index $i$ is a $M^5$ one,  we may define the corresponding contravariant components $P^i=-P_i$ and similarly for $Q_i^{(\rho)}$, as in Refs. \cite{CCC,CdS1}. These operators allow us to introduce the basis  $\{H, P_i,Q^{(\rho)}_i,J^{(\rho)}_i\}$ with the new commutators
\begin{eqnarray}
&&\left[ H, P_i \right]=i\omega P_i\,,\quad \hspace*{8.5 mm} \left[
H, Q^{(\rho)}_i
\right]=-i\omega Q^{(\rho)}_i\,,\label{HPQ}\\
&&\left[J^{(\rho)}_i , P_j \right]=i\varepsilon_{ijk} P_k\,,\quad
\left[ J^{(\rho)}_i, Q^{(\rho)}_j
\right]=i\varepsilon_{ijk} Q^{(\rho)}_k\,,\label{PJP}\\
&&\left[Q_i^{(\rho)}, P_j \right]=2 i \omega \delta_{ij} H + 2 i
\omega^2 \varepsilon_{ijk} J^{(\rho)}_k \,,\\
&&\left[Q_i^{(\rho)}, Q_j^{(\rho)} \right]=[P_i,P_j]=0\,.
\end{eqnarray}
Another basis is of the Poincar\' e type being formed by  $\{H,
P_i,J^{(\rho)}_i,K^{(\rho)}_i\}$. This has the commutation rules
given by Eqs. (\ref{ALG1}a), (\ref{ALG2}a), (\ref{ALG3}a),
(\ref{HPQ}a), (\ref{PJP}a) and
$\left[K^{(\rho)}_j, P_i\right]=i\delta_{ij}H-i\omega\varepsilon_{ijk}J^{(\rho)}_k$, while the commutator
(\ref{HHKR}b) has to be rewritten as $\left[ 
K_i^{(\rho)},H\right]=i P_i-i\omega K_i^{(\rho)}$.

The last two bases bring together the conserved energy (\ref{Ham})
and momentum (\ref{Pi}a) which are the only genuine orbital
operators, independent on $\rho$. What is specific for the de Sitter
symmetry is that these operators cannot be put simultaneously in
diagonal form since, according to Eq. (\ref{HPQ}a), they do not
commute with each other. 

When the covariant field transform under isometries  according to the CR  (\ref{Tx}) then the basis-generators transform as skew-symmetric tensors on $M^5$. Thus  the transformation rule (\ref{TxX}) can be rewritten as
\begin{equation}\label{TxXX}
T^{(\rho)}_{\frak g} X_{(AB)}^{(\rho)}{T_{\frak g}^{(\rho)}}^{-1}=\frak{g}_{A\,\cdot}^{\cdot\,C}{\frak g}_{B\,\cdot}^{\cdot\,D} X_{(CD)}^{(\rho)}\,,
\end{equation} 
where  the functions $\phi_{\frak g}$ result from the system  (\ref{zgz}) while ${\frak g}_{A\,\cdot}^{\cdot\,B}$ are the matrix elements of the matrix $\overline{\frak g}=\eta^5 {\frak g}\eta^5$. Hereby we conclude that such transformations mix among themselves all the basis-generators since there we do not have an invariant subgroup. This is in contrast with the flat case where the energy and momentum transform only among themselves  preserving the mass condition. For this reason the de Sitter invariants are more complicated involving simultaneously all the basis-generators,  as we shall show in the next section.

\subsection{Casimir operators and CR-UIR equivalence}

The first invariant of the CR $T^{(\rho)}$ is the quadratic Casimir
operator
\begin{eqnarray}
C^{(\rho)}_1&=&-\,\omega^2
\frac{1}{2}\,X_{(AB)}^{(\rho)}X^{(\rho)\,(AB)}\label{Q11}\\
&=&H^2-\omega^2({{\bf J}^{(\rho)}}\cdot
{{\bf J}^{(\rho)}}+{{\bf R}^{(\rho)}}\cdot
{{\bf R}^{(\rho)}}-{{\bf K}^{(\rho)}}\cdot{{\bf K}^{(\rho)}})\label{C14}\\
&=&H^2+3 i \omega H-{{\bf Q}^{(\rho)}}\cdot{{\bf P}}
-\omega^2{{\bf J}^{(\rho)}}\cdot{{\bf J}^{(\rho)}}\,.\label{C13}
\end{eqnarray}
which can be calculated according to Eqs.
(\ref{Ji})-(\ref{Ham}) and (\ref{Pi}). After a few manipulation we
obtain its definitive expression
\begin{equation}\label{CES}
C_1^{(\rho)}={\cal E}_{KG}+2i\omega e^{-\omega
t}S_{0i}^{(\rho)}\partial_i-\omega^2({{\bf S}}^{(\rho)})^2\,,
\end{equation}
depending on the Klein-Gordon operator of the scalar field,
\begin{equation}\label{EKG}
{\cal E}_{KG}= -\partial_t^2-3\,\omega\partial_t+e^{-2\omega
t}\Delta\,, \quad\Delta={{\bf \partial}\,}^2\,.
\end{equation}

The second Casimir operator,
\begin{equation}\label{Q2a}
C^{(\rho)}_2=-\eta^5_{AB}W^{(\rho)\, A}W^{(\rho)\,B}\,,
\end{equation}
is written with the help of the five-dimensional vector-operator
$W^{(\rho)}$ whose components read \cite{Gaz}
\begin{equation}\label{WW}
W^{(\rho)\, A}=\frac{1}{8}\,\omega\, \varepsilon^{ABCDE}
X_{(BC)}^{(\rho)}X_{(DE)}^{(\rho)}\,,
\end{equation}
where $\varepsilon^{01234}=1$ and the factor $\omega$  assures the
correct flat limit.  After a little calculation we obtain the
concrete form of these components,
\begin{eqnarray}
W^{(\rho)}_0&=&\,\omega\, {\bf J}^{(\rho)}\cdot{\bf R}^{(\rho)}
\,,\label{WW0}\\
{W}^{(\rho)}_i&=&H\,{J}_i^{(\rho)}+\omega\,
\varepsilon_{ijk}{K}^{(\rho)}_j {R}_k^{(\rho)}\,,\\
W^{(\rho)}_4&=&\,-\omega\,
{\bf J}^{(\rho)}\cdot{\bf K}^{(\rho)}\,,\label{WW5}
\end{eqnarray}
which indicate that $W^{(\rho)}$  plays an important role in
theories with spin, similar to that of the Pauli-Lubanski operator
(\ref{PaLu}) of the Poincar\' e symmetry. For example, the helicity
operator is now $W^{(\rho)}_0-W^{(\rho)}_4={S}^{(\rho)}_i {P}_i$.

Replacing then the components (\ref{WW0})-(\ref{WW5}) in Eq.
(\ref{Q2a}) we are faced with a complicated calculation but which can
be performed using algebraic codes on computer. Thus we obtain the
closed form of the second Casimir operator \cite{CCC},
\begin{eqnarray}
C_2^{(\rho)}&=& -\omega^2({\bf S}^{(\rho)})^2(t^2\partial_t^2-2t\partial_t + 2) +
2 \omega^2 t^2(iS_{0k}^{(\rho)}-\varepsilon_{ijk}S_i^{(\rho)}
S_{0j}^{(\rho)})\partial_k\partial_t\nonumber\\
&&+{\omega t}\left[({\bf S_0}^{(\rho)})^2\Delta
-(S_i^{(\rho)}S_j^{(\rho)}+S_{0i}^{(\rho)}S_{0j}^{(\rho)})
\partial_i\partial_j\right]\nonumber\\
&&-2i\omega^2  t( S_i^{(\rho)} S_k^{(\rho)}
S_{0i}^{(\rho)}+S_{0k}^{(\rho)})\partial_k\,.\label{Q221}
\end{eqnarray}

In the case of fields with unique spin $s$  we must select the reps. $\rho_s=(s,0)\oplus(0,s)$, for which we have to replace $S^{(s,0)}_{0i}=- i S^{(s)}_{i}$ and  $S^{(0,s)}_{0i}= i S^{(s)}_{i}$ in Eq. (\ref{Q221}) finding the remarkable identity
\begin{equation}\label{Q1Q2}
C_2^{\rho_s}=C_1^{\rho_s}{\bf S}^2-2\omega^2
{\bf S}^2+\omega^2 ({\bf S}^2)^2\,
\end{equation}
where we denote simply ${\bf S}={\rm diag}({\bf S}^{(s)}, {\bf S}^{(s)})$. 

Furthermore, it is interesting to look for the invariants of a particle at rest in the chart $\{t,{\bf x}\}$. This has the vanishing momentum ($P_i\sim 0$) so that $H$ acts as $i\partial_t$ and, therefore, it can be put in diagonal form its eigenvalue representing  just the rest
energiy $E_0$. Then, for each subspace  ${\cal V}_s \subset{\cal
V}_{(\rho)}$ of a given spin $s$, we obtain the eigenvalues of the
first Casimir operator,
\begin{equation}\label{rest}
C_1^{(\rho)}\sim E_0^2+3i\omega E_0 -\omega^2 s(s+1)\,,
\end{equation}
using Eqs. (\ref{CES}) and (\ref{EKG}) while those of the
second Casimir operator,
\begin{equation}\label{rest2}
C_2^{(\rho)}\sim s(s+1)(E_0^2+3i\omega E_0 -2 \omega^2)\,,
\end{equation}
result from Eq. (\ref{Q221}). These eigenvalues are real numbers
so that the rest energies, $E_0=\Re E_0-\frac{3i\omega}{2}$, must be
complex numbers whose imaginary parts are due to the decay produced
by the de Sitter expansion. 

The above results demonstrate that the induced CRs are reducible to direct sums of UIRs of the principal series \cite{linrep}, $(p,q)$, whose weights satisfy  \cite{CCC}
\begin{equation}\label{weights}
p=s\,,\quad q(1-q)=\frac{1}{\omega^2}\,(\Re E_0)^2+\frac{1}{4}\,.
\end{equation}
What is new here is that we meet only one type of UIRs, from the principal series, which are completely determined by the rest energy and the spin defined as in special relativity.

Finally, we specify that the physical interpretation adopted here is
correct since in the flat limit we recover the Poincar\' e generators \cite{CCC}.  We observe first that the
generators (\ref{Ji}) are independent on $\omega$ having the same
form as in the Minkowski case, $J_k^{(\rho)}=\hat J_k^{(\rho)}$. The
other generators have the limits
\begin{equation}
\lim_{\omega\to 0}H=\hat H=i\partial_t\,,\quad \lim_{\omega\to
0}(\omega R_i^{(\rho)})=\hat P_i=i\partial_i\,, \quad
\lim_{\omega\to 0}K_i^{(\rho)}=\hat K_i^{(\rho)} \,,
\end{equation}
which means that the basis $\{ H, P_i, J^{(\rho)}_i, K^{(\rho)}_i\}$
of the algebra $s(M)=sp(2,2)$ tends to the basis $\{\hat H, \hat
P_i,\hat J^{(\rho)}_i,\hat K^{(\rho)}_i\}$ of the $s(M_0)$ algebra
when $\omega\to 0$. Moreover, the Pauli-Lubanski operator
(\ref{PaLu}) is the flat limit of the five-dimensional
vector-operator (\ref{WW}) since
\begin{equation}
\lim_{\omega\to 0}W^{(\rho)}_0=\hat W^{(\rho)}_0\,,\quad
\lim_{\omega\to 0}W^{(\rho)}_i=\hat W^{(\rho)}_i\,, \quad
\lim_{\omega\to 0}W^{(\rho)}_4=0 \,.
\end{equation}
Under such circumstances the limits of our invariants read
\begin{equation}
\lim_{\omega\to 0}C_1^{(\rho)}=\hat C_1=\hat P^2\,,
\quad \lim_{\omega\to 0}C_2^{(\rho)}=\hat {\cal
C}_2^{(\rho)}\,,
\end{equation}
indicating that their physical meaning may be related to the mass
and spin of the matter fields in a similar manner as in special
relativity.

\section{The Dirac field on de Sitter spacetimes}

In the absence of a complete rep. theory like the Wigner one,  we must study the 
CR-UIR equivalence resorting to covariant field equations able to give us the structure of the covariant fields. Then, bearing in mind that the de Sitter UIRs are well-studied \cite{linrep,linrep1}, we can establish the CR-UIR equivalence by studying the CR and UIR Casimir operators in configuration and momentum reps..

In what follows we concentrate on the Dirac equation on the de Sitter spacetime since this is the only field equation on this background giving the natural rest energy $\Re E_0=m$ \cite{CCC}.

\subsection{Invariants of the spinor CR}

Let us consider the action of the free Dirac field $\psi$ of mass $m$ in the frame $\{t,{\bf x};e\}$  \cite{CPF},
\begin{equation}\label{action}
{\cal S}[e,\psi]=\int\, d^{4}x\sqrt{g}\left\{
\frac{i}{2}[\overline{\psi}\gamma^{\hat\alpha}D_{\hat\alpha}\psi-
(\overline{D_{\hat\alpha}\psi})\gamma^{\hat\alpha}\psi] -
m\overline{\psi}\psi\right\}\,,
\end{equation}
where  $g=|\det(g_{\mu\nu})|=(\omega t)^{-4}$ and $\overline\psi=\psi^+\gamma^0$. This action defines the Lagrangian theory of the Dirac field that yields the relativistic scalar product 
\begin{equation}\label{sp}
\left< \psi,\psi^{\prime}\right>=\int_{D}\frac{d^{3}x}{(-\omega t)^{3}}\, 
\overline{\psi}(x)\gamma^{0}\psi^{\prime}(x)  \,,
\end{equation}
and the Dirac equation
\begin{equation}\label{EDir}
({\cal E}_D-m)\psi(x)=\left[-i\omega t\left(\gamma^0\partial_{t}+\gamma^i\partial_i\right)
+\frac{3i\omega}{2}\gamma^{0}-m\right]\psi(x)=0\,,
\end{equation}
that  can be analytically solved either in momentum or energy bases with correct orthonormalization and completeness properties \cite{CPF,CSB}. 

The general solution of this equation in the spin-momentum rep. \cite{CSB},
\begin{equation}\label{psiab}
\psi(t,{\bf x})=\int d^3 p \sum_{\sigma}\left[U_{{\bf p},\sigma}(x)
a({\bf p},\sigma)+V_{{\bf p}, \sigma}(x){b}^{\dagger}({\bf p},
\sigma) \right]\,,
\end{equation}
is written in terms of the field operators, $a$ and $b$ (satisfying canonical anti-commutation rules), and  the particle and antiparticle fundamental spinors of momentum ${\bf p}=(p^1,p^2,p^3)$ (with $p=|{\bf p}|$) and polarization $\sigma=\pm\frac{1}{2}$, 
\begin{equation}
U_{{\bf p},\sigma}(t,{\bf x}\,)=\frac{1}{(2\pi)^{\frac{3}{2}}}\, u_{{\bf p},\sigma}(t) e^{i{\bf p}\cdot{\bf x}}\,,\quad
V_{{\bf p},\sigma}(t,{\bf x}\,)=\frac{1}{(2\pi)^{\frac{3}{2}}}\, v_{{\bf p},\sigma}(t) e^{-i{\bf p}\cdot{\bf x}}
\end{equation}
whose time-dependent terms have the form \cite{CSB,CQED}
\begin{eqnarray}
u_{{\bf p},\sigma}(t)&=&  \frac{i}{2}\left(\frac{\pi p}{\omega}\right)^{\frac{1}{2}}(\omega t)^2\left(
\begin{array}{r}
e^{\frac{1}{2}\pi\mu}H^{(1)}_{\nu_{-}}(-p t) \,
\xi_{\sigma}\\
e^{-\frac{1}{2}\pi\mu}H^{(1)}_{\nu_{+}}(-p t) \,\frac{{\bf \sigma}\cdot{\bf p}}{p}\,\xi_{\sigma}
\end{array}\right)\,,
\label{Ups}\\
v_{{\bf p},\sigma}(t)&=&\frac{i}{2} \left(\frac{\pi p}{\omega}\right)^{\frac{1}{2}}(\omega t)^2 \left(
\begin{array}{r}
e^{-\frac{1}{2}\pi\mu}H^{(2)}_{\nu_{-}}(-p t)\,\frac{{\bf \sigma}\cdot{\bf p}}{p}\,
\eta_{\sigma}\\
e^{\frac{1}{2}\pi\mu}H^{(2)}_{\nu_{+}}(-p t) \,\eta_{\sigma}
\end{array}\right)
\,,\label{Vps}
\end{eqnarray}
in the standard rep. of the Dirac matrices (with diagonal $\gamma^0$) and a fixed vacuum of the Bounch-Davies type \cite{CQED}. Obviously, the notation $\sigma_i$ stands for the Pauli matrices while  the point-independent Pauli spinors $\xi_{\sigma}$ and $\eta_{\sigma}= i\sigma_2 (\xi_{\sigma})^{*}$ are  normalized as $\xi^+_{\sigma}\xi_{\sigma'}=\eta^+_{\sigma}\eta_{\sigma'}=\delta_{\sigma\sigma'}$ \cite{CSB}. The terms giving the time modulation depend on the Hankel functions $H_{\nu_{\pm}}^{(1,2)}$   of indices 
\begin{equation}
\nu_{\pm}=\frac{1}{2}\pm i\mu \,, \quad  \mu=\frac{m}{\omega}\,,  
\end{equation}
whose properties are given in Appendix C.

Based on these properties we deduce the identities
\begin{equation}
{u}^+_{{\bf p},\sigma}(t){u}_{{\bf p},\sigma}(t)={v}^+_{{\bf p},\sigma}(t){v}_{{\bf p},\sigma}(t)=(-\omega t)^3
\end{equation}
allowing us to derive  the ortonormalization relations \cite{CPF} 
\begin{eqnarray}
&&\left<U_{{\bf p},\sigma},U_{{\bf p}^{\,\prime},\sigma^{\prime}}\right>=
\left<V_{{\bf p},\sigma},V_{{\bf p}^{\,\prime},\sigma^{\prime}}\right>=
\delta_{\sigma\sigma^{\prime}}\delta^3 ({\bf p}-{\bf p}^{\,\prime})\,,
\label{orto1}\\
&&\left<U_{{\bf p},\sigma},V_{{\bf p}^{\,\prime},\sigma^{\prime}}\right>=
\left<V_{{\bf p},\sigma},U_{{\bf p}^{\,\prime},\sigma^{\prime}}\right>=
0\,,\label{orto2}
\end{eqnarray}
that yield  the useful  inversion formulas, 
\begin{equation}\label{invers}
a({\bf p},\sigma)=\left<U_{{\bf p},\sigma},\psi\right>\,, \quad 
b({\bf p},\sigma)=\left<\psi,V_{{\bf p},\sigma}\right>\,.
\end{equation}
 Moreover, it is not hard to verify that these spinors are charge-conjugated to each other,
\begin{equation}\label{conj}
V_{{\bf p},\sigma}=(U_{{\bf p},\sigma})^{c}=C
(\overline{U}_{{\bf p},\sigma})^T \,, \quad C=i\gamma^2\gamma^0\,,
\end{equation}
and represent a {\em complete} system of solutions in the sense that \cite{CPF}
\begin{equation}\label{compl}
\int d^3 p \sum_{\sigma}\left[
U_{{\bf p},\sigma}(t,{\bf x})U^{+}_{{\bf p},\sigma}(t,{\bf x}^{\,\prime})+
V_{{\bf p},\sigma}(t,{\bf x})V^{+}_{{\bf p},\sigma}(t,{\bf x}^{\,\prime})
\right]=e^{-3\omega t}\delta^3 ({\bf x}-{\bf x}^{\,\prime})\,.
\end{equation}

The Dirac field transforms under isometries $x\to x'=\phi_{\frak g}(x)$ (with ${\frak g}\in I(M)$) according to the CR  $T_{\frak g} : \psi(x) \to (T_{\frak g}\psi)(x')=A_{\frak g}(x)\psi(x)$ induced by $\rho_D$ whose generators have the forms  (\ref{Ham}) - (\ref{Ri}).   Then, according to  Eqs.  (\ref{CES}) and (\ref{EDir}) we obtain the identity
\begin{equation}\label{Q1E}
C_1= {\cal E}_D^2+\frac{3}{2}\,\omega^2{\bf
1}_{4\times4} \sim m^2+\frac{3}{2}\,\omega^2\,,
\end{equation}
giving the first Casimir operator whose index $({\rho_D})$ is omitted since here we consider $\rho_D$ as the fundamental rep. of the $SL(2,{\Bbb C})$ group.
This result and Eq. (\ref{rest}) yield the rest energy of the
Dirac field,
\begin{equation}\label{EnD}
E_0=-\frac{3i\omega}{2}\pm m\,,
\end{equation}
which has a natural simple form where the decay (first) term is
added to the usual rest energy of special relativity. A similar
result can be obtained by solving the Dirac equation with vanishing
momentum.

The second invariant results from Eqs. (\ref{Q1Q2}) and
(\ref{Q1E}) if we take into account that
$({\bf S})^2=\frac{3}{4}\,{\bf 1}_{4\times4}$. Thus we
find
\begin{equation}\label{Q2E}
C_2=\frac{3}{4}\,{\cal E}_D^2+\frac{3}{16}\,
\omega^2{\bf 1}_{4\times4} \sim
\frac{3}{4}\left(m^2+\frac{1}{4}\,\omega^2\right)=\omega^2
s(s+1)\nu_+\nu_-\,,
\end{equation}
where  $s=\frac{1}{2}$ is the spin and $\nu_{\pm}=\frac{1}{2}\pm
i\frac{m}{\omega}$ are the indices of the Hankel functions giving
the time modulation of the Dirac spinors of the momentum basis
\cite{CPF}.

These invariants define the UIRs  that in the flat limit become Wigner's  UIRs  since 
\begin{equation}
\lim_{\omega\to 0} C_1\sim m^2\,,\quad
\lim_{\omega\to 0} C_2\sim \frac{3}{4}\,m^2\,,
\end{equation}
while the rest energy takes the value predicted in special relativity, $E_0 \to \pm m$. 

\subsection{Invariants of UIRs in momentum representation} 

The  inversion formulas (\ref{invers}) allow us to write the transformation rules in momentum rep. as
\begin{eqnarray}
(T_{\frak g}a)({\bf p},\sigma)&=&\left<U_{{\bf p},\sigma},[\rho_s(A_{\frak g})\psi]\circ \phi^{-1}_{\frak g}\right> \nonumber\\
&=&\int d^3p'\sum_{\sigma'}\left<U_{{\bf p},\sigma},[\rho_s(A_{\frak g})U_{{{\bf p}\,}',\sigma'}]\circ \phi^{-1}_{\frak g}\right> a({{\bf p}\,}',\sigma')\,, \label{Tg}\\
 (T_{\frak g}b)({\bf p},\sigma)&=&\left<[\rho_s(A_{\frak g})\psi]\circ \phi^{-1}_{\frak g},V_{{\bf p},\sigma}\right>\nonumber \\
 &=&\int d^3p'\sum_{\sigma'} \left<[\rho_s(A_{\frak g})V_{{{\bf p}\,}',\sigma'}]\circ \phi^{-1}_{\frak g},V_{{\bf p},\sigma}\right>  b({{\bf p}\,}',\sigma')\,,   \label{Tg1}
\end{eqnarray}
but, unfortunately, these scalar products are complicated integrals that cannot be solved in the general case. 

Nevertheless, one can prove that for the particular isometries  ${\frak g}={\frak g}({\bf \omega}){\frak g}({\bf a})$ of the Euclidean subgruop $E(3)$, formed by a rotation  ${\frak g}({\bf \omega})\in SO(3)$ and a translation ${\frak g}({\bf a})\in T(3)_P$,  we find the linear transformations, 
\begin{equation}
(T_{\frak g}a)({\bf p},\sigma)=D^{(\frac{1}{2})}_{\sigma \sigma'}({\bf \omega}) a({{\bf p}\,}',\sigma') e^{i {\bf a}\cdot{{\bf p}\,}'}\,, \quad {{\bf p}\,}'=R^{-1}({\bf \omega}){\bf p}\,,
\end{equation}
that  are somewhat similar with those of the Wigner theory. However,  there are isometries that cannot be brought in a such simple form as, for example,  those of the adjoint Abelian subgroup $T(3)_Q$. Therefore, we must abandon the group transformations focusing on the generators of the corresponding Lie algebras in momentum rep..

Any self-adjoint generator $X=\overline{X}$ of the spinor rep. of the $s(M)$ algebra gives rise to a {conserved} one-particle operator of the QFT, 
\begin{eqnarray}
{\cal X}&=&:\left<\psi, X\psi\right>:={\cal X}^{(+)}+{\cal X}^{(-)}\nonumber\\
&=&\int d^3 p\, 
\left[\alpha^{\dagger}({\bf p}) {\tilde X}^{(+)}\alpha({\bf p})
+\beta^{\dagger}({\bf p})\tilde X^{(-)} \beta({\bf p})\right]\,,\label{X}
\end{eqnarray}
calculated respecting the normal ordering of the operator products \cite{BDR}.  These operators are Hermitian with respect to the scalar product of the state space, ${\cal X}={\cal X}^{\dagger}$, since $X$ is self-adjoint. The operators ${\tilde X}^{(\pm)}$ are the generators of CRs in momentum rep. acting on the operator valued Pauli spinors, 
\begin{equation}
\alpha({\bf p})=\left(\begin{array}{l}
a({\bf p},\frac{1}{2})\\
a({\bf p},-\frac{1}{2})
\end{array}\right)\,,\quad
\beta({\bf p})=\left(\begin{array}{l}
b({\bf p},\frac{1}{2})\\
b({\bf p},-\frac{1}{2})
\end{array}\right) \,.
\end{equation}
As observed in Ref. \cite{Nach}, the straightforward method for finding the structure of these operators is to evaluate the entire expression  (\ref{X}) by using the form  (\ref{psiab}) where the field operators $a$ and $b$ satisfy the {canonical} anti-commutation rules \cite{Nach,CPF}.

For this purpose we consider several identities written with the notation $\partial_{p^i}=\frac{\partial}{\partial p^i}$ as 
\begin{eqnarray}
&& H\,U_{{\bf p},\sigma}(t,{\bf x})=-i\omega
\left(p^i\partial_{p^{i}}+\frac{3}{2}\right)
U_{{\bf p},\sigma}(t,{\bf x})\,,\nonumber\\
&& H\,V_{{\bf p},\sigma}(t,{\bf x})=-i\omega
\left(p^i\partial_{p^{i}}+\frac{3}{2}\right)
V_{{\bf p},\sigma}(t,{\bf x})\,,\nonumber
\end{eqnarray}
that help us to eliminate some multiplicative operators and the time derivative when we  inverse the Fourier transform. Furthermore, by applying the Green theorem and calculating on computer terms of the form ${u}^+_{{\bf p},\sigma}(t)F(t,{\bf p}){u}_{{\bf p},\sigma}(t)$,  ${v}^+_{{\bf p},\sigma}(t)F(t,{\bf p}){v}_{{\bf p},\sigma}(t)$,..., etc., we find two  {\em identical} reps. whose basis generators read,   ${\tilde P}^{(\pm)\,i}=\tilde P^i=p^i$ and
\begin{eqnarray}
{\tilde H}^{(\pm)}&=&\omega {\tilde X}^{(\pm)}_{(04)}=i \omega \left (p^i \partial_{p^i}+\frac{3}{2}\right)\,,\label{Hamp}\\
{\tilde J_i}^{(\pm)}&=& \frac{1}{2}\,\varepsilon_{ijk}
{\tilde X}^{(\pm)}_{(jk)}=-i\varepsilon_{ijk}p^j\partial_{p^k}+\frac{1}{2}\sigma_i\,,\\
{\tilde K}^{(\pm)}_i&=&{\tilde X}^{(\pm)}_{(0i)}= i {\tilde H}^{(\pm)}\partial_{p^i}+\frac{\omega}{2}p^i\Delta_p - p^i \frac{{p}^2+m^2}{2\omega {p}^2}\nonumber\\
&&~~~~~~~~~~+\frac{1}{2}\varepsilon_{ijk}\left(i\omega \partial_{p^j}- p^j\frac{m}{{ p}^2}\right)\sigma_k\,,\\
{\tilde R}^{(\pm)}_i&=&{\tilde X}^{(\pm)}_{(i4)}=-{\tilde K}^{(\pm)}_i-\frac{1}{\omega}\, p^i\,,\label{Rip}
\end{eqnarray}
where $p=|{\bf p}|$ and $\Delta_p=\partial_{p^i}\partial_{p^i}$.
These basis generators satisfy the specific  $sp(2,2)$ commutation rules of the form (\ref{ALG1})-(\ref{HHKR}). Moreover, it is not difficult to verify that these are Hermitian operators with respect to the scalar products of the momentum rep.
\begin{equation}
\left<\alpha, \alpha'\right>=
\int d^3 p\, 
\alpha^{\dagger}({\bf p}) \tilde \alpha({\bf p})\,, \quad 
\langle \beta,\beta'\rangle=\int  d^3p \beta^{\dagger}({\bf p})\tilde \beta({\bf p})\,.
\end{equation}
Therefore, we can conclude that these operators generate a pair of  {\em unitary} reps. of the  group $S(M)$. Note that the operator (\ref{Hamp}) was obtained by Nachtmann many years ago \cite{Nach} while the other generators were derived recently in Ref. \cite{Cnew}.

Since all the UIRs of the group $S(M)$ are classified \cite{linrep},  we can study  the equivalence and reducibility of these reps. simply  by calculating the Casimir operators in momentum rep.. By using the same definitions as  in the configuration rep. we write the first  Casimir operator as,
\begin{equation}
\tilde C_1=-\,
\frac{1}{2}\,\omega^2 \tilde X_{(AB)}\tilde X^{(AB)}\label{Q1b}\,,
\end{equation}
while the second one,
\begin{equation}\label{Q2b}
\tilde C_2=-\eta^5_{AB}\tilde W^{A}\tilde W^{B}\,,\quad \tilde W^{A}=\frac{1}{8}\, \omega \varepsilon^{ABCDE}
\tilde X_{(BC)}\tilde X_{(DE)}\,,
\end{equation}
is written in terms of the Pauli-Lubanski operator of components $\tilde W^A$.   After performing the calculation on computer we find \cite{Cnew}
\begin{eqnarray}
\tilde W_0^{(\pm)}&=&\frac{\omega}{4}({\bf \sigma}\cdot{\bf p})\Delta_p+\frac{\omega\nu_{-}}{2}{\bf \sigma}\cdot {\bf \partial}_p+\frac{i m}{2 p^2}({\bf \sigma}\cdot {\bf p})\, {\bf p}\cdot{\bf \partial}_p\nonumber\\
&&+\frac{m^2-{p}^2+ 2i\omega m}{4 p^2\omega}{\bf \sigma}\cdot{\bf p}\,,\label{W0}\\
\tilde W_i^{(\pm)}&=&\frac{i}{2}({\bf \sigma}\cdot{\bf p}) \partial_{p^i}+\frac{i\nu_{-}}{2}\sigma_i - \frac{m}{2\omega p^2}\,({\bf \sigma}\cdot{\bf p}) p^i\,, \\
\tilde W_4^{(\pm)}&=&\tilde W_0^{(\pm)}+\frac{1}{2 \omega}{\bf \sigma}\cdot{\bf p}\,.\label{W4}
\end{eqnarray}
With these preparation we obtain the Casimir operators (\ref{Q1b}) and (\ref{Q2b}) as 
\begin{eqnarray}
\tilde C_1^{(\pm)}&=&\omega^2[-s(s+1)-(q+1)(q-2)]=  m^2+\frac{3\omega^2}{2}\,,\label{Cp1} \\
\tilde C_2^{(\pm)}&=&\omega^2[-s(s+1)q(q-1)]\nonumber\\
&=&\omega^2
s(s+1)\nu_+\nu_-=
\frac{3}{4}\left(m^2+\frac{\omega^2}{4}\right)\,,\label{Cp2}
\end{eqnarray}
recovering thus the Casimir eigenvalues (\ref{Q1E}) and (\ref{Q2E}) obtained in configuration rep.. 

\subsection{CR-UIR equivalence in QFT}

The above result shows  that the  identical spinor reps. we obtained here are  UIRs of the principal series corresponding to the canonical weights $(s,q)$ with  $s=\frac{1}{2}$ and $q=\nu_{\pm}$.  In other words the spinor CR of the Dirac theory is equivalent with the orthogonal sum of the equivalent UIRs of the particle and antiparticle sectors.  This suggests that the UIRs  $(s, \nu_{\pm})$ of the group $S(M)=Sp(2,2)$ can be seen as being analogous to the Wigner ones  of the Dirac theory in Minkowski spacetime. Thus we find that the particle and antiparticle UIRs are equivalent which means that the particles and antiparticles are identical from the point of view of the external symmetry, just as in special relativity.

Here we adopted phase factors such that these equivalent UIRs are in fact identical, having the same generators in momentum prep., $\tilde X^{(+)}_{(AB)}= \tilde X^{(-)}_{(AB)}$.
However, in general, these equivalent spinor UIRs may not coincide since  the expressions of their basis generators are strongly dependent on the arbitrary phase factors of the fundamental spinors whether these depend on ${\bf p}$. Thus if we change   
\begin{equation}\label{gaugeUV}
U_{{\bf p},\sigma}\to e^{i\chi^+({\bf p})} U_{{\bf p},\sigma}\,,\quad
V_{{\bf p},\sigma}\to e^{-i\chi^-({\bf p})} V_{{\bf p},\sigma}\,,
\end{equation}
with $\chi^{\pm}({\bf p})\in \Bbb R$, performing simultaneously  the associated transformations,
\begin{equation}\label{gaugeab}
\alpha({\bf p})\to e^{-i\chi^+({\bf p})}\alpha({\bf p})\,,\quad
\beta({\bf p})\to e^{-i\chi^-({\bf p})}\beta({\bf p})\,,
\end{equation}
that preserves the form of $\psi$, we find that the operators ${\tilde P}^i=p^i$   keep their forms while the other generators are changing since in all the above formulas we must replace
\begin{equation}
\partial_{p^i} \to\partial_{p^i} -i \partial_{p^i}\chi^{\pm}({\bf p})\,.
\end{equation} 
For example,  the Hamiltonian operators become, 
\begin{equation}
{\tilde H}^{(\pm)}\to  {\tilde H}^{(\pm)} + \omega p^i \partial_{p^i}\chi^{\pm}({\bf p})\,.
\end{equation}
Obviously,  these transformations are nothing other than unitary transformations among equivalent UIRs.  Note that thanks to this mechanism one can fix  suitable phases for determining desired forms of the basis generators, keeping thus under control the flat and rest limits of these operators in the scalar \cite{Nach,BDurr,Cscalar} or  Dirac \cite{Nach,CSB}  field theory on $M$.

At the level of  QFT, the operators $\{{\cal X}_{(AB)}\}$  of the form (\ref{X}), corresponding to the differential operators (\ref{Hamp}) - (\ref{Rip}), 
\begin{eqnarray}\label{Xfin}
{\cal X}_{(AB)}&=&:\langle \psi,X_{(AB)}\psi\rangle:  ={\cal X}^{(+)}_{(AB)}+{\cal X}^{(-)}_{(AB)}\nonumber\\
&=&\int d^3 p\, 
\left[\alpha^{\dagger}({\bf p}) {\tilde X}^{(+)}_{(AB)}\alpha({\bf p})
+\beta^{\dagger}({\bf p})\tilde X^{(-)}_{(AB)} \beta({\bf p})\right]\,,
\end{eqnarray}
generate a reducible operator valued CR which can be decomposed as the orthogonal sum of  CRs, generated by $\{{\cal X}_{(AB)}^{(+)}\}$ and $\{{\cal X}_{(AB)}^{(-)}\}$,  that are equivalent between themselves and equivalent with the UIR $(\frac{1}{2},\nu_{\pm})$ of the $sp(2,2)$ algebra. These one-particle operators are the principal conserved quantities of the Dirac theory corresponding to the de Sitter isometries via Noether theorem. According to Eqs. (\ref{algXX}) we can write their action on the field operators in configuration or momentum reps. as
\begin{eqnarray}
\left[{\cal X}_{(AB)}, \psi(x)\right]&=&-(X_{(AB)}\psi)(x)\,, \\
\left[{\cal X}_{(AB)}^{(+)}, \alpha({\bf p})\right]&=&-(\tilde X_{(AB)}^{(+)}\alpha)({\bf p})\,, \\
\left[{\cal X}_{(AB)}^{(-)}, \beta({\bf p})\right]&=&-(\tilde X_{(AB)}^{(-)}\beta)({\bf p})\,. 
\end{eqnarray}
The transformation of the field operators under isometries ${\frak g}={\frak g}(\xi)\in I(M)$ is  performed by the {\em unitary} operators 
\begin{equation}\label{Unit}
{\cal U}_{\frak g}=\exp\left(i \xi ^{AB}{\cal X}_{(AB)}\right)
\end{equation}
which have the action 
\begin{equation}
{\cal U}_{\frak g}\psi(x){\cal U}_{\frak g}^{\dagger}=\left(T_{\frak{g}}\psi\right)(x)=A_{\frak g}[\phi^{-1}_{\frak{g}}(x)] \psi [\phi^{-1}_{\frak{g}}(x)]\,,
\end{equation}
where $T_{\frak g}$ is the rep. (\ref{Tx})  induced by $\rho_D$. In momentum rep. we find the corresponding transformations 
\begin{equation}
{\cal U}_{\frak g}a({\bf p},\sigma){\cal U}_{\frak g}^{\dagger}=\left(T_{\frak{g}} a\right)({\bf p},\sigma)\,, \quad 
{\cal U}_{\frak g}b({\bf p},\sigma){\cal U}_{\frak g}^{\dagger}=\left(T_{\frak{g}} b\right)({\bf p},\sigma)\,,
\end{equation}
given by Eqs. (\ref{Tg}) and (\ref{Tg1}). Finally, we derive the transformation rule of the generators (\ref{Xfin}) that yields
\begin{eqnarray}
{\cal U}_{\frak g}^{\dagger}{\cal X}_{(AB)}{\cal U}_{\frak g}&=&:\langle T_{\frak g}^{-1} \psi,X_{(AB)}T_{\frak g}^{-1}\psi\rangle: \nonumber\\
&=&:\langle \psi,T_{\frak g}X_{(AB)}T_{\frak g}^{-1}\psi\rangle:={\frak g}_{A\,\cdot}^{\cdot\,C}{\frak g}_{B\,\cdot}^{\cdot\,D}{\cal X}_{(CD)}\,,
\end{eqnarray}
as it results from Eq. (\ref{TxXX}). Hereby, we observe again that one of the advantages of the second quantization is of encapsulating all the transformation properties under isometries into the unique unitary  operator (\ref{Unit}).

It remains to study the invariants of the QFT. In the relativistic quantum mechanics  the conserved observables  form an algebra freely generated by the basis-generators where we find the operators $W_A$ or the Casimir ones derived by using operator multiplication. However, this method does not hold in QFT since the product of two operators  of the form (\ref{X}) gives terms which are no longer one-particle operators. Therefore, we must build the one-particle operators by using exclusively the rule (\ref{X}). We define thus  the Pauli -Lubanski operator,
\begin{eqnarray}
{\cal W}_A&=&:\langle \psi,W_A\psi\rangle:  ={\cal W}^{(+)}_A+{\cal W}^{(-)}_A\nonumber\\
&=&\int d^3 p\, 
\left[\alpha^{\dagger}({\bf p}) {\tilde W}^{(+)}_A\alpha({\bf p})
+\beta^{\dagger}({\bf p})\tilde W^{(-)}_A \beta({\bf p})\right]\,,
\end{eqnarray}
according to  Eqs. (\ref{W0})-(\ref{W4}), and the Casimir operators
\begin{eqnarray}
{\cal C}_1&=&:\langle \psi,C_1\psi\rangle:=\left (\mu^2+\frac{3}{2}\right){\cal N}\,,\\
{\cal C}_2&=& :\langle \psi,C_2\psi\rangle:=\frac{3}{4}\left(\mu^2+\frac{1}{4}\right){\cal N}\,,
\end{eqnarray}
where
\begin{equation}
 {\cal N}={\cal N}^{(+)}+{\cal N}^{(-)}=\int d^3 p\, 
\left[\alpha^{\dagger}({\bf p}) \alpha({\bf p})
+\beta^{\dagger}({\bf p}) \beta({\bf p})\right]
\end{equation}
is the usual operator of the total number of particles and antiparticles resulted from Eqs. (\ref{Cp1}) and (\ref{Cp2}). 

It is remarkable that  the particle and antiparticle sectors bring similar contributions to the conserved quantities corresponding to isometries. Thus we can say that these quantities are {\em additive}, e. g., the energy of a many particle system is the sum of the individual energies of particles and antiparticles.  This additivity holds for the entire theory of the spacetime symmetries in contrast with the conserved charges of the internal symmetries that take different values for particles and antiparticles. For example, the charge operator  corresponding to the $U(1)_{em}$ gauge symmetry \cite{CQED}  reads ${\cal Q}=q :\langle \psi,\psi\rangle: =q({\cal N}^{(+)}-{\cal N}^{(-)})$.

The principal conclusion is that the spinor CR is equivalent with a direct sum of a pair of equivalent UIRs of the principal series, $(\frac{1}{2},\nu_{\pm})$, whose weights (\ref{weights}) correspond to the  rest energy $\Re E_0=m$  which is the same as in special relativity. We recovered thus the general result of section 4 completed with the rest energy given by the Dirac equation. 

\section{Concluding remarks}

Here we have seen that the QFT on the de Sitter background has similar features as in the flat case. Thus the covariant quantum fields transform according to CRs induced by the reps. of the group $\hat G=SL(2,\Bbb C)$ that are equivalent with orthogonal sums of UIRs of the group $S(M)=Sp(2,2)$ whose  specific invariants depend only on particle masses and spins. The example is the  spinor CR of the Dirac theory that is induced by the linear rep. $(\frac{1}{2},0)\otimes (0,\frac{1}{2})$ of the group $\hat G$ but is equivalent to the orthogonal sum of two equivalent UIRs of the group $S(M)$ labelled by $(\frac{1}{2},\nu_{\pm})$. Thus, at least in the case of the Dirac field, we recover a similar conjuncture as in  the Wigner theory of the induced reps. of the Poincar\' e group in special relativity. However, the principal difference is that the transformations of  the Wigner UIRs  can be written in closed forms while in our case this cannot be done because of the technical difficulties  in solving the integrals (\ref{Tg}) and (\ref{Tg1}). For this reason we were forced to restrict ourselves to study only the reps.  of the corresponding algebras.

This is not an impediment since physically speaking we are interested  to know the properties of the basis generators (in configuration or momentum rep.) since these give rise to the conserved observables (i. e. the one-particle operators) of QFT, associated to the de Sitter isometries. It is remarkable that the particle and antiparticle terms of these operators bring additive contributions since  the particle and antiparticle operators transform alike under isometries,  just as in special relativity. Notice that this result was obtained by Nachtmann \cite{Nach} for the scalar UIRs and partially for the spinor ones.   Now  this property is demonstrated completely in the Dirac case such that we can conclude that all the one-particle operators corresponding to the de Sitter isometries are additive,  regardless the spin. In fact, we verified thus that the connection between spin and statistic is universal, having the same effects on curved backgrounds just as in the flat case. 

The principal  problem that remains unsolved here is how to build a Wigner type theory on the de Sitter manifolds  able to define the structure of the covariant fields without  using field equations. This means to solve first the problem of the UIR transformations in momentum rep. and then  to look for a general definition of  mass or even of a mass operator on $M$, related to the Casimir operators of the UIRs of the $S(M)$ group.  We note that despite of the well-known classical results \cite{Nach,BDurr}  it remains a discrepancy between the manners in which the masses of bosons and fermions depend on the de Sitter invariants \cite{CCC}. We hope that the results presented here and our recent de Sitter relativity \cite{CdS1} will offer one new tools in solving these delicate problems.

\appendix
\section{Finite-dimensional  representations of the $sl(2,{\Bbb C})$ algebra}

The standard basis of the $sl(2,{\Bbb C})$ algebra is formed by the generators 
${\bf J}=(J_1,J_2,J_3)$ and ${\bf K}=(K_1,K_2,K_3)$ that satisfy \cite{Nai,WKT,BARA}
\begin{equation}
[J_i,J_j]=i\varepsilon_{ijk}J_k\,,\quad [J_i,K_j]=i\varepsilon_{ijk}K_k\,, \quad
[K_i,K_j]=-i\varepsilon_{ijk}J_k\,,
\end{equation} 
having the Casimir operators $c_1=i{\bf J}\cdot {\bf K}$ and $c_2={\bf J}^2-{\bf K}^2$. The linear combinations $A_i=\frac{1}{2}\left(J_i +i K_i\right)$ and $B_i=\frac{1}{2}\left(J_i-i K_i\right)$ form two independent $su(2)$ algebras satisfying
\begin{equation}
[A_i,A_j]=i\varepsilon_{ijk}A_k\,,\quad [B_i,B_j]=i\varepsilon_{ijk}B_k\,, \quad
[A_i,B_j]=0\,.
\end{equation} 
Consequently, any finite-dimensional irrep.  $\tau=(j_1, j_2)$ is carried by the space ${\cal V}_{\tau}={\cal V}_{j_1}\otimes {\cal V}_{j_2}$ of the direct product $(j_1)\otimes (j_2)$ of the UIRs $(j_1)$  and $(j_2)$ of the $su(2)$ algebras $(A_i)$ and respectively $(B_i)$. These irreps. are labeled either by the $su(2)$ weights $(j_1,j_2)$ or giving the values of the Casimir operators $c_1=j_1(j_1+1)-j_2(j_2+1)$ and $c_2=2[j_1(j_1+1)+j_2(j_2+1)]$.

The fundamental reps. defining the $sl(2,{\Bbb C})$ algebra are either the irrep. 
$(\frac{1}{2},0)$ generated by $\{\frac{1}{2}\sigma_i, -\frac{i}{2}\sigma_i\}$ or the irrep. $(0, \frac{1}{2})$ whose generators are $\{\frac{1}{2}\sigma_i, \frac{i}{2}\sigma_i\}$. Their direct sum form the spinor  rep. $\rho_D= (\frac{1}{2},0)\oplus (0, \frac{1}{2})$ of the Dirac theory. In applications it is convenient to consider $\rho_D$ as the fundamental rep. since this is the simplest rep. with unique spin $s=\frac{1}{2}$  in which one can define an invariant form by using the Dirac conjugation.

This conjecture can be generalized easily considering pairs of {\em adjoint} irreps.,   $\tau=(j_1,j_2)$ and  $\dot \tau=(j_2,j_1)$, which have the same spin content while  their generators are related as ${\bf J}^{(\dot\tau)}={\bf J}^{(\tau)}$ and  ${\bf K}^{(\dot\tau)}=-{\bf K}^{(\tau)}$. Since  the operators ${\bf A}$ and ${\bf B}$ are Hermitian, generating  UIRs of the $su(2)$ algebra,  we have ${\bf J}^+={\bf J}$ and  ${\bf K}^+=-{\bf K}$  such that we can write
\begin{equation}
({\bf J}^{(\tau)})^+={\bf J}^{(\tau)}\,, \quad ({\bf K}^{(\tau)})^+={\bf K}^{(\dot\tau)}\,.
\end{equation} 
Hereby we conclude that the invariant forms can be constructed only when we use {\em symmetric} reps. $\rho=\cdots \tau_1\oplus\tau_2\cdots \dot \tau_1\oplus\dot\tau_2\cdots$
containing only pairs of adjoint reps. and self-adjoint irreps. $\tau=\dot\tau=(j,j)$. Then the matrix $\gamma_{(\rho)}$ may be constructed having the matrix elements 
\begin{equation}
\langle \tau_1,s_1\sigma_1|\gamma_{(\rho)}|\tau_2,s_2\sigma_2\rangle=\delta_{\tau_1\dot\tau_2}\delta_{s_1s_2}\delta_{\sigma_1\sigma_2}\,.
\end{equation}
Moreover, for such reps. we can construct at any time the charge conjugation matrix ${C}_{(\rho)}$ having the matrix elements 
\begin{equation}
\langle \tau_1,s_1\sigma_1|C_{(\rho)}|\tau_2,s_2\sigma_2\rangle=\eta(\tau_1)\delta_{\tau_1\dot\tau_2}\delta_{s_1s_2}(-1)^{s_1-\sigma_1}\delta_{\sigma_1,-\sigma_2}\,.
\end{equation}
and acting as  
\begin{equation}\label{Cconj}
\rho(S_{\hat\alpha\hat\beta})^*=-{C}_{(\rho)}\rho(S_{\hat\alpha\hat\beta}){C}_{(\rho)}^{-1}~~ \to ~~\rho(A)^*={C}_{(\rho)}\rho(A){C}_{(\rho)}^{-1}\,,
\end{equation}
where $^*$ denotes the complex conjugation. The phases $\eta(\tau)=\pm1$ have to be chosen in order to accomplish the orthogonality relation (\ref{orto2a}).

Finally we note that the canonical basis $\{|\tau, j\lambda\rangle\}$ defines the {\em chiral} rep. while a new basis in which $\gamma_{(\rho)}$ becomes diagonal gives the so called standard rep.. This terminology comes from the Dirac theory where $\gamma_{(\rho_D)}=\gamma^0$ and all the Dirac matrices  may have such types of reps. \cite{Th}.

\section{Dirac matrices}

The Dirac matrices with local indices, $\gamma^{\hat\mu}$, satisfy  $\{\gamma^{\hat\mu},\gamma^{\hat\nu}\}=2\eta^{\hat\mu\hat\nu}I$ and give rise to the $sl(2,{\Bbb C})$ generators according to Eq. (\ref{Sgg}). There are two usual reps. of the Dirac matrices, the  
standard rep., 
\begin{equation}
\gamma^0=\left(
\begin{array}{cc}
1&0\\
0&-1
\end{array}\right)\,,\quad
\gamma^i=\left(
\begin{array}{cc}
0&\sigma_i\\
-\sigma_i&0
\end{array}
\right)\,,\quad
\gamma^5=\left(\begin{array}{cc}
0&1\\
1&0
\end{array}\right)\,,
\end{equation}
and the chiral one
\begin{equation}
\gamma^0=\left(
\begin{array}{cc}
0&1\\
1&0
\end{array}\right)\,,\quad
\gamma^i=\left(
\begin{array}{cc}
0&\sigma_i\\
-\sigma_i&0
\end{array}
\right)\,,\quad
\gamma^5=\left(\begin{array}{cc}
-1&0\\
0&1
\end{array}\right)\,,
\end{equation}
where the $sl(2,{\Bbb C})$ generators are reducible to the subspaces of the irreps. $(\frac{1}{2},0)$ and $(0,\frac{1}{2})$ of $\rho_D$ \cite{WKT},
\begin{equation}
S_i=\frac{1}{2}\left(
\begin{array}{cc}
\sigma_i&0\\
0&\sigma_i
\end{array}\right)\,,\quad
S_{0i}=\frac{1}{2}\left(
\begin{array}{cc}
-i\sigma_i&0\\
0&i\sigma_i
\end{array}\right)\,.
\end{equation}

Notice that the space ${\cal V}_D$ of the rep. $\rho_D$ can be seen as the space of a fundamental representation of the $su(2,2)$ Lie algebra \cite{TY} since  the $SL(2,{\Bbb C})$ generators, $S^{\hat\mu\hat\nu}$, the Dirac matrices $\gamma^{\hat\mu}$,  $\gamma^5$, and $\gamma^{\hat\mu}\gamma^5$  satisfy the commutation rules defining the $su(2,2)$ algebra, 
\begin{eqnarray}
&[\gamma^{\hat\nu},\gamma^{5}]=2\gamma^{\hat\nu}\gamma^5\,,\quad~~&~~ [\gamma^{\hat\nu}\gamma^5,\gamma^{5}]=2\gamma^{\hat\nu}\nonumber\\
&[\gamma^{\hat\mu},\gamma^{\hat\nu}]=-4iS^{\hat\mu\hat\nu}\,,\quad&~~~[S^{\hat\mu\hat\nu},\gamma^{5}]\,=\,0\,, \nonumber\\
&[\gamma^{\hat\mu},\gamma^{\hat\nu}\gamma^5]=2\eta^{\hat\mu\hat\nu}\gamma^5\,,\quad~~~
&~~~[S^{\hat\mu\hat\nu},\gamma^{\hat\sigma}]=i(\eta^{\hat\nu\hat\sigma}\gamma^{\hat\mu}
-\eta^{\hat\mu\hat\sigma}\gamma^{\hat\nu})\,,\nonumber \\
&[\gamma^{\hat\mu}\gamma^5,\gamma^{\hat\nu}\gamma^5]=4iS^{\hat\mu\hat\nu}\,, \quad
~~~~~~~~&{[}S^{\hat\mu\hat\nu},\gamma^{\hat\sigma}\gamma^5{]}
=i(\eta^{\hat\nu\hat\sigma}\gamma^{\hat\mu}\gamma^5
-\eta^{\hat\mu\hat\sigma}\gamma^{\hat\nu}\gamma^5)\,,\nonumber
\end{eqnarray}
\begin{equation}\label{comSS1}
{[}S^{\hat\mu\hat\nu},S^{\hat\sigma\hat\tau}{]}=i(
\eta^{\hat\mu\hat\tau}S^{\hat\nu\hat\sigma}-\eta^{\hat\mu\hat\sigma}S^{\hat\nu\hat\tau}+\eta^{\hat\nu\hat\sigma}S^{\hat\mu\hat\tau}
-\eta^{\hat\nu\hat\tau}S^{\hat\mu\hat\sigma})\,.
\end{equation}
Moreover, these matrices can be considered as generating even a superalgebra since they satisfy the closed anti-commutation relations
\begin{eqnarray}
&\{\gamma^{\hat\nu},\gamma^{5}\}=0\,,\quad~~~~~~~~~~~~& ~~\{\gamma^{\hat\nu}\gamma^5,\gamma^{5}\}=0 \nonumber\\
&\{\gamma^{\hat\mu},\gamma^{\hat\nu}\}=2\eta^{\hat\mu\hat\nu}I\,,\quad~~~~~~~
&~~~\{S^{\hat\mu\hat\nu},\gamma^{5}\}=-i\,\tilde\varepsilon\,^{\hat\mu\hat\nu\,\cdot\,\cdot}
_{\,\cdot\,\cdot\,\hat\sigma\hat\tau}\,S^{\hat\sigma\hat\tau}\,,\nonumber\\
&~~~~\{\gamma^{\hat\mu},\gamma^{\hat\nu}\gamma^5\}=-2\,\tilde\varepsilon\,^{\hat\mu\hat\nu\,\cdot\,\cdot}
_{\,\cdot\,\cdot\,\hat\sigma\hat\tau}\,S^{\hat\sigma\hat\tau}\,,\quad~~~~
&~~~\{S^{\hat\mu\hat\nu},\gamma^{\hat\sigma}\}=\tilde\varepsilon\,^{\hat\mu\hat\nu\hat\sigma\,\cdot}
_{\,\cdot\,\cdot\,\cdot\,\,\hat\tau}
\,\gamma^{\hat\tau}\gamma^5\,,\nonumber\\
&\{\gamma^{\hat\mu}\gamma^5,\gamma^{\hat\nu}\gamma^5\}=-2\eta^{\hat\mu\hat\nu}I\,,\quad ~~~~~~~~~~
&\{S^{\hat\mu\hat\nu},\gamma^{\hat\sigma}\gamma^5\}
=\tilde\varepsilon\,^{\hat\mu\hat\nu\hat\sigma\,\cdot}_{\,\cdot\,\cdot\,\cdot\,\,\hat\tau}
\,\gamma^{\hat\tau}\,,\nonumber
\end{eqnarray}
\begin{equation}
\{S^{\hat\mu\hat\nu},S^{\hat\sigma\hat\tau}\}=\frac{1}{2}\,
(\eta^{\hat\mu\hat\sigma}\eta^{\hat\nu\hat\tau}-\eta^{\hat\nu\hat\sigma}\eta^{\hat\mu\hat\tau})I-\frac{i}{2}\,
\tilde\varepsilon\,^{\hat\mu\hat\nu\hat\sigma\hat\tau}\gamma^5\,.\nonumber 
\end{equation}
Here we denoted by $I$  the identity operator on ${\cal V}_D$ and 
$\varepsilon\,^{\hat\mu\hat\nu\hat\sigma\hat\tau}$ the usual Levi-Civita symbol \cite{MTW}
adopting the convention $\varepsilon_{0123}=-\varepsilon^{0123}=1$. In addition we have identities \cite{MTW}
\begin{equation}
\varepsilon\,_{\hat\mu\hat\nu\hat\sigma\hat\tau}\,\varepsilon\,^{\hat\alpha\hat\beta\hat\sigma\hat\tau}=
-2\left(\delta^{\hat\alpha}_{\hat\mu}\delta^{\hat\beta}_{\hat\nu}-\delta^{\hat\alpha}_{\hat\nu}\delta^{\hat\beta}_{\hat\mu}
\right)\,, \quad
\varepsilon\,_{\hat\mu\hat\nu\hat\sigma\hat\tau}\,\varepsilon\,^{\hat\alpha\hat\nu\hat\sigma\hat\tau}
=-6\,\delta^{\hat\alpha}_{\hat\mu}\,,
\end{equation}
that can be used in current calculations.

\section{Some properties of Hankel functions}

According to the general properties of the Hankel functions \cite{GR}, we
deduce that those used here, $H^{(1,2)}_{\nu_{\pm}}(z)$, with
$\nu_{\pm}=\frac{1}{2}\pm i \mu$ and $z\in \Bbb R$, are related among
themselves through $[H^{(1,2)}_{\nu_{\pm}}(z)]^{*}
=H^{(2,1)}_{\nu_{\mp}}(z)$ and satisfy the  identities
\begin{equation}\label{H3}
e^{\pm \pi k}
H^{(1)}_{\nu_{\mp}}(z)H^{(2)}_{\nu_{\pm}}(z)
+ e^{\mp \pi k} H^{(1)}_{\nu_{\pm}}(z)H^{(2)}_{\nu_{\mp}}(z)=\frac{4}{\pi z}\,.
\end{equation}

\section*{Acknowledgments}

This work was partially supported by a grant of the Ministry of National Education and Scientific Research, RDI Programme for Space Technology and Advanced Research - STAR, project number 181/20.07.2017.


\end{document}